\documentclass[onecolumn]{aa}
\usepackage{graphicx}
\usepackage{txfonts}
\usepackage{url}

\begin{document}

\title{``TNOs are Cool'': A survey of the trans-Neptunian region}
\subtitle{{\bf X.} Analysis of classical Kuiper belt objects from \emph{Herschel}\thanks{{\it Herschel}
is an ESA space observatory with science instruments provided by
European-led Principal Investigator consortia and with important participation from NASA.} and
\emph{Spitzer} observations}

\author{E. Vilenius\inst{1} \and C. Kiss\inst{2} \and T. M\"uller\inst{1} \and M. Mommert\inst{3,4} \and
P. Santos-Sanz\inst{5,6} \and A. P\'al\inst{2} \and
J. Stansberry\inst{7} \and M. Mueller\inst{8,9} \and N. Peixinho\inst{10,11} \and
E. Lellouch\inst{6} \and S. Fornasier\inst{6,12}  
\and A. Delsanti\inst{6,13} \and
A. Thirouin\inst{5} \and J.~L.~Ortiz\inst{5} \and R. Duffard\inst{5} 
\and
D. Perna\inst{6} 
\and F. Henry\inst{6} 
}

\institute{Max-Planck-Institut f\"ur extraterrestrische Physik, Postfach 1312, Giessenbachstr., 85741 Garching, Germany\\
\email{vilenius@mpe.mpg.de}
\and
Konkoly Observatory, Research Centre for Astronomy and Earth Sciences, Konkoly Thege 15-17, H-1121 Budapest, Hungary
\and
Deutsches Zentrum f\"ur Luft- und Raumfahrt e.V., Institute of Planetary Research, Rutherfordstr. 2, 12489 Berlin, Germany
\and
Northern Arizona University, Department of Physics and Astronomy, PO Box 6010, Flagstaff, AZ, 86011, USA
\and
Instituto de Astrof\'isica de Andaluc\'ia (CSIC), Glorieta de la Astronom\'ia s/n, 18008-Granada, Spain
\and
LESIA-Observatoire de Paris, CNRS, UPMC Univ. Paris 06, Univ. Paris-Diderot, France
\and
Stewart Observatory, The University of Arizona, Tucson AZ 85721, USA
\and
SRON Netherlands Institute for Space Research, Postbus 800, 9700 AV Groningen, the Netherlands
\and
UNS-CNRS-Observatoire de la C{\^o}te d'Azur, Laboratoire Cassiope\'e, BP 4229, 06304 Nice Cedex 04, France
\and
Center for Geophysics of the University of Coimbra, Geophysical and
Astronomical Observatory of the University of Coimbra, Almas de Freire, 3040-004 Coimbra, Portugal
\and
Unidad de Astronom\'{\i}a, Facultad de Ciencias B\'asicas, Universidad de
Antofagasta, Avenida Angamos 601, Antofagasta, Chile
\and
Univ. Paris Diderot, Sorbonne Paris Cit\'{e}, 4 rue Elsa Morante, 75205 Paris, France
\and
Laboratoire d'Astrophysique de Marseille, CNRS \& Universit\'e de Provence, 38 rue Fr\'ed\'eric Joliot-Curie,
13388 Marseille Cedex 13, France
}

\date{Received July 31, 2013; accepted January 19, 2014}

\abstract
{The Kuiper belt is formed of planetesimals which failed to grow to planets and its dynamical structure
has been affected by Neptune. The classical Kuiper belt contains objects both from a low-inclination,
presumably primordial, distribution and from a high-inclination dynamically excited population.}
{Based on a sample of classical TNOs with observations at thermal wavelengths we determine radiometric sizes,
geometric albedos and thermal beaming factors for each object as well as 
study sample properties
of dynamically hot and cold classicals.}
{Observations near the thermal peak of TNOs using infra-red space
telescopes are combined with optical magnitudes
using the radiometric technique with near-Earth asteroid thermal model (NEATM).
We have determined three-band flux
densities from \emph{Herschel}/PACS observations at 70.0, 100.0 and $160.0\ \mathrm{\mu m}$
and \emph{Spitzer}/MIPS at 23.68 and $71.42\ \mathrm{\mu m}$ when available.
We use reexamined absolute visual magnitudes  
from the literature and ground based programs in support of \emph{Herschel} observations.
}
{We have analysed 18 classical TNOs with previously unpublished data and re-analysed
previously published targets with updated data reduction to
determine their sizes and geometric albedos as well as beaming factors when data quality allows. 
We have combined these samples with classical TNOs with radiometric results in the
literature for the analysis of sample
properties 
of a total of 44 
objects. We find  
a median geometric albedo for 
cold classical TNOs of 
$0.14_{-0.07}^{+0.09}$
and for dynamically hot classical TNOs, 
excluding the Haumea family and dwarf planets, $0.085_{-0.045}^{+0.084}$.
We have determined the bulk densities of Borasisi-Pabu ($2.1_{-1.2}^{+2.6}$ g cm$^{-3}$),
Varda-Ilmar\"e
% 2003 MW$_{12}$ 
($1.25_{-0.43}^{+0.40}$ g cm$^{-3}$) and 2001 QC$_{298}$ ($1.14_{-0.30}^{+0.34}$ g cm$^{-3}$) as well as
updated previous density estimates of four targets. We have determined the slope
parameter of the 
debiased
cumulative size distribution of dynamically hot classical TNOs
as $q$=2.3$\pm$0.1 in the diameter range 100$<$$D$$<$500 km. For dynamically cold classical TNOs
we determine $q$=5.1$\pm$1.1 in the diameter range 160$<$$D$$<$280~km as the
cold classical TNOs have a smaller maximum size.
}
{}


\maketitle

\section{Introduction}
Transneptunian objects (TNO) 
are  
believed, based on theoretical modeling,
to represent the leftovers from the formation
process of the solar system.
Different classes of objects may probe different regions
of the protoplanetary disk and provide clues
of different ways of accretion in those regions (\cite{Morbidelli2008}). Basic physical
properties of TNOs, such as size and albedo, have been challenging
to measure. Only a few brightest TNOs have size estimates using
direct optical imaging  
(e.g. Quaoar with \emph{Hubble}; \cite[2004]{Brown2004}). Stellar occultations by TNOs provide
a possibility to obtain an accurate size estimate, but these events
are rare and require a global network of observers (e.g. Pluto's moon Charon by \cite{Sicardy2006};
and a member of the dynamical class of classical TNOs, 2002 TX$_{300}$, by
\cite{Elliot2010}, 2010).  
Predictions of
occultations are limited by astrometric uncertainties of both TNOs
and stars. Combining observations of reflected light at optical wavelengths
with thermal emission data, which for TNOs
peaks in the far-infrared wavelengths, allows us to determine both
size and geometric albedo for large samples of targets. This {\it radiometric method}
using space-based ISO (e.g. \cite{Thomas2000}), \emph{Spitzer}
(e.g. \cite[2008]{Stansberry2008}, \cite[2009]{Brucker2009}) 
and \emph{Herschel} data (\cite[2010]{Muller2010}, \cite[2010]{Lellouch2010},
\cite[2010]{Lim2010}, \cite[2012]{SantosSanz2012},
\cite[2012]{Mommert2012}, \cite[2012]{Vilenius2012}, \cite{Pal2012}, \cite[2013]{Fornasier2013})
has already  changed the size estimates of several TNOs compared
to those obtained by using an assumed albedo and has revealed a large
scatter in albedos and differences
between dynamical classes of TNOs.

Observations at thermal wavelengths also provide
information about thermal properties (\cite[2013]{Lellouch2013}).
Depending on the thermal or thermophysical
model selected it is possible to derive the thermal beaming factor
or the thermal inertia, and constrain other surface properties.
Ground-based submillimeter observations can also be used to determine
TNO sizes using the radiometric method (e.g. \cite{Jewitt2001b}), but
this technique has been limited to very few targets so far. 

TNOs, also known as Kuiper belt objects (KBO), have diverse dynamical 
properties and they are divided into classes.
Slightly different definitions and names for these classes are available
in the literature. Classical TNOs (hereafter CKBO) reside mostly beyond Neptune
on orbits which are not very eccentric and not in mean motion resonance 
with Neptune.
We use the \cite{Gladman2008}
classification:
CKBOs are non-resonant TNOs which do not belong to any other TNO class. The eccentricity
limit is $e$\,$\lesssim$\,$0.24$, beyond which objects belong to {\it detached objects}
or {\it scattering/scattered objects}. Classical TNOs with semimajor axis
$39.4$\,$<$\,$a$\,$<$\,$48.4\ \mathrm{AU}$ occupy the {\it main classical belt}, whereas
{\it inner} and {\it outer} classicals exist at smaller and larger semi-major axis, respectively.
Apart from the Gladman system, another common classification is defined by
the Deep Eplictic Survey Team~(DES, \cite{Elliot2005} 2005).
For the work presented here, the most notable difference between the
two systems is noticed with
high-inclination objects. Many of them are not CKBOs in the DES system.

In the inclination/eccentricity space CKBOs show two different populations, which
have different frequency of binary systems (\cite{Noll2008}), 
different luminosity functions (LF; \cite{Fraser2010} 2010), different average 
geometric albedos (\cite{Grundy2005, Brucker2009} 2009) and different color distributions
(\cite[2008]{Peixinho2008}).
The low-inclination ``cold'' classicals are limited to the main classical belt and
have a higher average albedo, more binaries and a steeper LF-derived size distribution
than high-inclination ``hot'' classicals. Some amount of transfer between the hot and cold CKBOs
is possible with an estimated maximum of 5\% of targets in either population originating from the other
than its current location (\cite{Volk2011}).

The \emph{``TNOs are Cool'': A survey of the trans-Neptunian region}
open time key program (\cite{Muller2009})
of \emph{Herschel Space Observatory} 
has observed
12 cold CKBOs, 29 hot CKBOs, and five CKBOs in the inner classical belt,
which are considered to be dynamically hot. In addition, 
eight CKBOs have
been observed only by \emph{Spitzer Space Telescope}, whose TNO sample was mostly
overlapping with the \emph{Herschel} one. 

This paper is organized in the following way. 
We begin by describing our target sample in Section~\ref{Targetsample},
followed by \emph{Herschel} observations and their planning in Section~\ref{Hobs}
and \emph{Herschel} data reduction in Section~\ref{dataredux}. More far-infrared data
by \emph{Spitzer} are presented in Section~\ref{Sobs} and absolute visual magnitudes in Section~\ref{auxobs}.
Thermal modeling combining the above mentioned data is described in
Section~\ref{model} and the results for targets in our sample in Section~\ref{resultsection},
comparing them with earlier results when available (Section~\ref{resultscomp}).
In Section~\ref{discussions} we discuss sample properties, cumulative size distributions, 
correlations 
and binaries 
as well as debiasing of the measured size distributions.
Conclusions of the sample analysis are given in Section~\ref{conclude}.

\section{Target sample and observations}
\subsection{Target sample}
\label{Targetsample}
The classification of targets in the ``TNOs are Cool'' program
within the \cite{Gladman2008} framework is based on the list used by
Minor Bodies in the Outer Solar System 2 data base (MBOSS-2, \cite{Hainaut2012} 2012 and
C. Ejeta, \emph{priv. comm.}). The inclination distributions of the dynamically
cold and hot components of CKBOs are partly overlapping.
A cut-off limit of $i$\,$=$\,$4.5\degr$ is used in this work, and the inclinations
we use from the Minor Planet Center are measured with respect
to the ecliptic plane, which deviates slightly from the invariable plane of the Solar System,
or the average Kuiper belt plane. All the cold CKBOs with measured sizes available
have inclinations $i$$<$4.0$\degr$ (see Table~\ref{collection_colds} in Section~\ref{discussions}).
Three CKBOs listed as dynamically hot in Table~\ref{collection} 
(2000 OK$_{67}$, 2001 QD$_{298}$ and Altjira)
have 4.5$<$$i$$<$5.5$\degr$. Since the two populations
overlap in the inclination space some targets close to the cut-off limit could belong
to the other population.
In the DES classification system all targets
in Table~\ref{table_overview} with $i$\,$>$\,$15\degr$ would belong to the scattered-extended
class of TNOs. DES uses the Tisserand parameter and
orbital elements in the CKBO/scattered objects distinction, whereas the Gladman system requires
an object to be heavily interacting with Neptune in order to be classified as a scattered object.

In this work we 
have reduced the flux densities 
of 16 CKBOs observed with \emph{Herschel}.
Together with \cite{Vilenius2012} (2012), \cite{Fornasier2013} (2013) and
\cite{Lellouch2013} (2013)
this work completes the set of CKBOs observed
by \emph{Herschel}, except for the classical Haumea family members with water
signatures in their spectra,
whose properties differ
from the ``bulk'' of CKBOs (Stansberry et al., {\it in prep.}). Photometric
3-band observations were done in 2010-2011 with \emph{Herschel}/PACS in the wavelength
range 60--$210\ \mathrm{\mu m}$. Seven of the 16 targets have been observed also with two
bands of \emph{Spitzer}/MIPS imaging photometer at 22--$80\ \mathrm{\mu m}$ in 2004-2008. 
In addition, our target sample (Table~\ref{table_overview}) includes two previously unpublished targets
2003 QR$_{91}$ and 2001 QC$_{298}$ observed only with MIPS and
are included in the radiometric analysis of this work.

The relative amount of binaries among the cold CKBOs with radiometric measurements is high
(Table 6) with only very few non-binaries. While the binary fraction
among cold CKBOs has been estimated to be 29\% (\cite{Noll2008}) the actual frequency may
be higher because there probably are binaries which have not been resolved with current
observing capabilities.
Furthermore, in the target selection process of "TNOs are Cool" we aimed to have a
significant sample of binary TNOs observed, and the highest binary fraction of all
dynamical classes is in the cold sub-population of CKBOs.

For sample analysis 
we have 
included all CKBOs with radiometric results
from this work and literature, some of which have been reanalyzed in this work.
We achieve a total sample size of 44 
targets
detected with either \emph{Herschel} 
or \emph{Spitzer} 
(Tables~\ref{collection_colds} and \ref{collection}). 
The absolute V-magnitudes ($H_{\mathrm{V}}$, see Section~\ref{auxobs}) of the combined sample
range from about 3.5 to 8.0 mag (0.1-8.0 mag if 
dwarf planets are included).
A typical characteristic of CKBOs is 
that bright classicals
have systematically higher inclinations than fainter ones
(\cite{Levison2001} 2001). Our combined sample
shows a moderate correlation (see Section~\ref{corr_other}) between
absolute magnitude and inclination at 4$\sigma$ level of significance. For 
about half of the targets a color taxonomy is available.
Almost all very red targets (RR) in the combined sample are at inclinations
$i$\,$<$\,$12\degr$. This is consistent with \cite{Peixinho2008} (2008) who report a color break
at $i$\,$=$\,$12\degr$ instead of at the cold/hot boundary inclination
near 5$\degr$.
\begin{table*}
\centering
\caption{Orbital and color properties of the sample of 18 classical TNOs with new flux densities presented in this work.}
\begin{tabular}{lcccccccll}
\hline
\hline
Target & & $q$ & $Q$  & $i$     & $e$ & $a$ & Color                 & Spectral slope\tablefootmark{d}  & V-R  \\
       & &(AU) & (AU) & (\degr) &     &     & taxa\tablefootmark{a} & (\% / 100 nm)                    &      \\
\hline
\object{(2001 QS$_{322}$)}                &                     & 42.3 & 46.1 &     0.2  &  0.043 & 44.2 & \ldots & \ldots                                  & \ldots  \\
\object{66652 Borasisi (1999 RZ$_{253}$)} & B                   & 40.0 & 47.8 &     0.6  &  0.088 & 43.9 & RR     & $33.8 \pm 2.7$\tablefootmark{e,f}  &  $0.646 \pm 0.058$\tablefootmark{f,l}   \\
\object{(2003 GH$_{55}$)}                 &                     & 40.6 & 47.3 &     1.1  &  0.076 & 44.0 & \ldots & $26.0 \pm 5.6$\tablefootmark{g}   &   $0.63 \pm 0.06$\tablefootmark{g}      \\
\object{135182 (2001 QT$_{322}$)} {\it in inner belt} &  & 36.6 & 37.9 &     1.8  &  0.018 & 37.2 & \ldots & $15.6 \pm 11.1$\tablefootmark{h}  &      $0.53 \pm 0.12$\tablefootmark{h} \\
\object{(2003 QA$_{91}$)}                 & B                   & 41.4 & 47.7 &     2.4  &  0.071 & 44.5 & \ldots & \ldots                            &  \ldots        \\
\object{(2003 QR$_{91}$)}                 & B                   & 38.1 & 55.0 &     3.5  &  0.182 & 46.6 & \ldots & \ldots                            &  \ldots        \\
\object{(2003 WU$_{188}$)}                & B                   & 42.4 & 46.3 &     3.8  &  0.043 & 44.3 & \ldots & \ldots                            &  \ldots        \\
\hline
\object{35671 (1998 SN$_{165}$)} {\it in inner belt} &   & 36.4 & 39.8 &     4.6  &  0.045 & 38.1 & BB                     & $6.9 \pm 3.1$\tablefootmark{f,i,j,k,l}  & $0.444 \pm 0.078$\tablefootmark{f,i,j,k,l}        \\
\object{(2001 QD$_{298}$)}                &                     & 40.3 & 45.1 &     5.0  &  0.056 & 42.7 & \ldots                 & $30.4 \pm 8.3$\tablefootmark{m}         & $0.67 \pm 0.09$\tablefootmark{m}         \\
\object{174567 Varda (2003 MW$_{12}$)}    & B                   & 39.0 & 52.2 &    21.5  &  0.144 & 45.6 & IR\tablefootmark{b,c}  & $19.2 \pm 0.6$\tablefootmark{n}         & \ldots           \\
\object{86177 (1999 RY$_{215}$)}          &                     & 34.5 & 56.5 &    22.2  &  0.241 & 45.5 & BR                     & $3.8 \pm 3.5$\tablefootmark{l,o,p}      & $0.358 \pm 0.090$\tablefootmark{l,o}       \\
\object{55565 (2002 AW$_{197}$)}          &                     & 41.2 & 53.2 &    24.4  &  0.127 & 47.2 & IR                     & $22.1 \pm 1.4$\tablefootmark{g,k,q,r,s} & $0.602 \pm 0.031$\tablefootmark{g,k,q,r,v} \\
\object{202421 (2005 UQ$_{513}$)}         &                     & 37.3 & 49.8 &    25.7  &  0.143 & 43.5 & \ldots                 & $18.1 \pm 2.0$\tablefootmark{t}         &  \ldots          \\
\object{(2004 PT$_{107}$)}                &                     & 38.2 & 43.1 &    26.1  &  0.060 & 40.6 & \ldots                 & \ldots                                  & $0.65 \pm 0.10$\tablefootmark{v}       \\
\object{(2002 GH$_{32}$)}                 &                     & 38.1 & 45.7 &    26.7  &  0.091 & 41.9 & \ldots                 & $24.8 \pm 4.7$\tablefootmark{u}         & $0.425 \pm 0.228$\tablefootmark{m,v,w} \\
\object{(2001 QC$_{298}$)}                & B                   & 40.6 & 52.1 &    30.6  &  0.124 & 46.3 & \ldots                 & $10.3 \pm 2.4$\tablefootmark{e,g,p}     & $0.490 \pm 0.030$\tablefootmark{g} \\
\object{(2004 NT$_{33}$)}                 &                     & 37.0 & 50.1 &    31.2  &  0.150 & 43.5 & BB-BR\tablefootmark{c} & \ldots                                  & \ldots           \\
\object{230965 (2004 XA$_{192}$)}         &                     & 35.5 & 59.4 &    38.1  &  0.252 & 47.4 & \ldots                 & \ldots                                  & \ldots           \\
\hline
\end{tabular}
\label{table_overview}
\tablefoot{Perihelion distance $q$, aphelion distance $Q$, inclination $i$, eccentricity $e$, semimajor axis $a$
(orbital elements from IAU Minor Planet Center, \url{http://www.minorplanetcenter.net/iau/lists/TNOs.html},
accessed June 2012), color taxonomy, spectral slope, and 
(V-R) color index ordered according to increasing inclination. 
The horizontal line marks the limit of dynamically
cold and hot classicals at $i$\,$=$\,$4.5\degr$ (Targets in the inner belt are dynamically hot
regardless of their inclination.). B denotes a known binary system (\cite{Noll2008},
except Varda (2003 MW$_{12}$) from \cite{Noll2009} and \cite{Benecchi2013}).
Targets are located in the main classical belt unless otherwise indicated.
\\
\\
{\bf References}.
\tablefoottext{a}{Taxonomic class from \cite{Fulchignoni2008} unless otherwise indicated.}
\tablefoottext{b}{\cite{Perna2010}.}
\tablefoottext{c}{\cite{Perna2013}.}
\tablefoottext{d}{Spectral slopes from MBOSS-2 online database (except 2005 UQ$_{513}$ and 2002 GH$_{32}$) of
\cite{Hainaut2012} (2012) at \url{http://www.eso.org/~ohainaut/MBOSS}, accessed October 2012.
References of original data indicated for each target.}
\tablefoottext{e}{\cite{Benecchi2009}.}
\tablefoottext{f}{\cite{Delsanti2001}.}
\tablefoottext{g}{\cite{Jewitt2007}.}
\tablefoottext{h}{\cite{Romanishin2010}.}
\tablefoottext{i}{\cite{Jewitt2001}.}
\tablefoottext{j}{\cite{Gilhutton2001}.}
\tablefoottext{k}{\cite{Fornasier2004}.}
\tablefoottext{l}{\cite{Doressoundiram2001}.}
\tablefoottext{m}{\cite{Doressoundiram2005a}.}
\tablefoottext{n}{\cite{Fornasier2009}.}
\tablefoottext{o}{\cite{Bohnhardt2002}.}
\tablefoottext{p}{\cite{Benecchi2011} (2011).}
\tablefoottext{q}{\cite{Doressoundiram2005b}.}
\tablefoottext{r}{\cite{DeMeo2009}.}
\tablefoottext{s}{\cite{Rabinowitz2007} (2007).}
\tablefoottext{t}{\cite{Pinilla2008}.}
\tablefoottext{u}{\cite{Carry2012}.}
\tablefoottext{v}{\cite{Snodgrass2010}.}
\tablefoottext{w}{\cite{SantosSanz2009} (2009).}
}
\end{table*}

\subsection{\emph{Herschel} observations}
\label{Hobs}
\emph{Herschel Space Observatory} (\cite{Pilbratt2010}) 
was orbiting the Lagrange 2 point of the Earth-Sun system in 2009-2013.
It has a 3.5 m radiatively cooled telescope and three science instruments inside
a superfluid helium cryostat. The photometer part of the PACS instrument (\cite{Poglitsch2010})
has a rectangular field of view of $1.75\arcmin$\,$\times$\,$3.5\arcmin$.
It has two bolometer arrays, the short-wavelength one
is for wavelengths 60\,--\,$85\ \mathrm{\mu m}$ or 85\,--\,$125\ \mathrm{\mu m}$,
selectable by a filter wheel, and the long-wavelength array for 125\,--\,$210\ \mathrm{\mu m}$.
The absolute calibration 1-$\sigma$ uncertainty is 5\% in all bands
(\cite{Balog2013}).
The detector pixel sizes are $3.2\arcsec \times 3.2\arcsec$
in the short-wavelength array, whereas the long-wavelength array has larger pixels of
$6.4\arcsec \times 6.4\arcsec$.
The instrument is continuously sampling the detectors and produces 40 frames/s, which are
averaged on-board by a factor of four. \emph{Herschel} recommended to use
the scanning technique for point sources instead of chopping and nodding, to achieve
better sensitivity (\cite{PACSrelnote}). Pixels in the image frames, 
sampled continuously while the telescope
was scanning, were mapped in the data reduction pipeline (see Section~\ref{dataredux})
into pixels of a sub-sampled output image.

Our observations (Table~\ref{table_obs}) with PACS followed the same strategy as in
\cite{Vilenius2012} (2012). We made three-band observations of all targets in two scan
directions of the rectangular array, and repeated the same observing sequence on a
second visit. We used mini-scan maps with 2-6 repetitions per observation.
The final maps are combinations of four observations/target, except at the
$160\ \mathrm{\mu m}$ band where all eight observations/target were available independent
of the filter wheel selection. To choose the number of repetitions, i.e. the duration
of observations, we used a thermal model (see Section~\ref{model}) to predict
flux densities. We adopted a default geometric albedo of 0.08 and a beaming factor of
1.25 for observation planning purposes. For two bright targets we used other values based
on earlier \emph{Spitzer} results (\cite[2008]{Stansberry2008}):
for 1998 SN$_{165}$ a lower geometric albedo of 0.04 and for 2002 AW$_{197}$ a
higher geometric albedo of 0.12. In the combined maps the predicted instrumental
signal-to-noise ratios (SNR) for the 16 targets with the above assumptions were
SNR$\sim$\,13 (faintest target SNR$\sim$\,4) at the 70 and $100\ \mathrm{\mu m}$ channels
and SNR$\sim$\,7 (faintest target SNR$\sim$\,2) at the $160\ \mathrm{\mu m}$ channel.
The sensitivity of the $70\ \mathrm{\mu m}$ channel is
usually limited by instrumental noise, while the aim of our combination of observations is to
remove the background confusion noise affecting the other two channels, most notably the
$160\ \mathrm{\mu m}$ band.

The selection of the observing window was optimized to utilize the lowest far-infrared confusion noise
circumstances~(\cite{Kiss2005}) of each target during the \emph{Herschel} mission.
Targets were visited twice within the same observing window with a similar set of
2x2 observations on each of the two 
visits for the purpose of background subtraction (\cite{Kiss2013}).
The time gap between the visits was 11-42 hours depending on the proper
motion of the target.

\subsection{PACS Data reduction}
\label{dataredux}
We used data reduction and image combination techniques developed within the
``TNOs are Cool'' key program
(\cite{Kiss2013} and references cited therein).
Herschel Interactive Processing Environment (HIPE\footnote{Data presented in this paper were
analysed using ``HIPE'', a joint development
by the Herschel Science Ground Segment Consortium, consisting of ESA, the 
NASA Herschel Science Center, and the HIFI, PACS and SPIRE consortia 
members, see \url{http://herschel.esac.esa.int/DpHipeContributors.shtml}.}, version 9.0 / CIB 2974)  
was used to produce Level 2 maps with modified scan map pipeline
scripts. 
The pipeline script provided a two-stage high-pass filtering procedure to handle the 1/f
noise, which is dominating the timelines of individual detectors in the PACS photometer arrays.
The script removes from each timeline, excluding the masked
parts of timelines where we expect the source to be present, a value obtained by a running median filter. The
filter width parameters we used were typically 8/9/16
readouts, and for some targets 10/15/25 readouts at the 70/100/$160\ \mathrm{\mu m}$
channels, respectively. We set the map-pixel sizes to 1.1\arcsec/pixel, 1.4\arcsec/pixel and
2.1\arcsec/pixel for the three channels, respectively, to properly sample the point spread
functions.

For combining the projected output images and reducing the background we use two
methods: ``super-sky-subtracted''
images (\cite[2009]{Brucker2009}, \cite[2012]{SantosSanz2012}) and ``double-differential'' images
(\cite[2012]{Mommert2012}, \cite{Kiss2013}). The ``super-sky'' is constructed by masking the source (or an area
surrounding the image center when the target is too faint to be recognized in individual images)
in each individual
image, combining these sky images and subtracting this combined background from each individual image.
Then, all background-subtracted images are co-added in the co-moving frame of the target.
The ``double-differential'' images are produced in a different way. Since the observing strategy
has been to make two sets of observations with similar settings, we subtract the combined images
of the two visits. This yields a positive and a negative beam of the moving source with
background structures eliminated. A duplicate of this image is shifted to match the positive beam
of the original image with the negative one of the duplicate. After subtracting these from
each other we have a double-differential image with one positive and two negative beams, where
photometry is done on the central, positive beam. It can be noted that this method works well
even if there is a systematic offset in target coordinates due to uncertain astrometry. A further
advantage is in the detection of faint sources: they should have one positive and two negative
beams in the final image
(with negative beams having half the flux density of the positive one).
In both methods of combining individual observations of a target we take into account the offsets and
uncertainties in pointing and assigned image coordinates
(\cite{Pal2012}, \cite{Kiss2013}).

Photometry is performed with DAOPHOT routines (\cite{Stetson1987}), 
which are available via commonly used astronomy software tools such as HIPE, IDL and IRAF (for details how
photometry is done in the ``TNOs are Cool'' program see \cite[2012]{SantosSanz2012}). 
 A color correction to flux densities
is needed because TNOs have a spectral energy distribution (SED) resembling
a black body whereas the PACS photometric
system assumes a flat SED. The correction, based on instrumental transmission and response
curves available from HIPE, is typically at the level of 2\% or less depending on the temperature of
the TNO. The color correction is fine-tuned in an iterative way (for details see \cite[2012]{Vilenius2012}).
For uncertainty estimation of the derived flux density we use 200 artificial implanted
sources within a region close to the source, excluding the target itself.

The color corrected flux densities from PACS are given in
Table~\ref{table_obs}, where also the absolute calibration uncertainty has been included in the
1-$\sigma$ error bars. The flux densities are preferably averaged from the photometry
results using the two techniques discussed above:
the ``super-sky-subtracted'' and the ``double-differential''. Since the super-sky-subtracted way
gives more non-detected bands than the double-differential way we take the average only
when the super-sky-subtracted method produces a 3-band detection, otherwise only
flux densities based on the double-differential images are used for a given target. In Table~\ref{table_obs}
the seven targets whose flux densities at $160\,\mathrm{\mu m}$ are $>$5 mJy
have flux densities averaged from the double-differential and super-sky-subtracted methods.

The flux density predictions used in the planning (Section~\ref{Hobs})
of these observations differ by factors of $\pm 2$ or more compared to the measured flux densities.
On the average, the measured values are 
lower ($\sim$ 50\%) than the predicted ones. Only three targets are
brighter than estimated in the PACS bands and there are four 
targets not detected in the PACS observations.
The average SNRs of detected targets are half of the average SNRs of the predictions used in
observation planning.

\begin{table*}
\centering
\caption{\emph{Herschel} observations and monochromatic flux densities at all three PACS bands. 2001 QY$_{297}$ and
Altjira from \cite{Vilenius2012} (2012) have been reanalysed in this work with changes in flux densities and radiometric results.}
\begin{tabular}{llrlcccrrr}
\hline
Target & 1st OBSIDs   & Dur.  & Mid-time & $r$  & $\Delta$ & $\alpha$ & \multicolumn{3}{c}{Flux densities (mJy)} \\
       & of visit 1/2 & (min) &          & (AU) & (AU)     & (\degr)  & $70\,\mathrm{\mu m}$ & $100\,\mathrm{\mu m}$ & $160\,\mathrm{\mu m}$ \\
\hline
2001 QS$_{322}$  & 1342212692/...2726 & 188.5 & 2011-Jan-15 22:54 & 42.36 & 42.78 &  1.22 & $1.8 \pm 1.1$                   &    $4.0 \pm 1.6$                  &  $4.3 \pm 2.0$  \\         
Borasisi         & 1342221733/...1806 & 226.1 & 2011-May-27 23:21 & 41.62 & 41.74 &  1.40 & $<$1.0                          &    $<$1.4                         &   $<$1.4 \\
2003 GH$_{55}$   & 1342212652/...2714 & 188.5 & 2011-Jan-15 13:14 & 40.84 & 41.16 &  1.31 & $2.0 \pm 1.0$                   &    $<$1.3                         &  $<$1.4 \\   
2001 QY$_{297}$  & 1342209492/...9650 & 194.8 & 2010-Nov-19 03:28 & 43.25 & 43.25 &  1.31 & $<$1.3                          &    $<$2.1                            & $<$2.1 \\
2001 QT$_{322}$  & 1342222436/...2485 & 226.1 & 2011-Jun-10 15:15 & 37.06 & 37.38 &  1.50 & $2.6 \pm 1.1$                   &    $<$6.7                         &   $<$1.5  \\
2003 QA$_{91}$   & 1342233581/...4252 & 226.1 & 2011-Dec-05 06:06 & 44.72 & 44.85 &  1.26 & $1.8 \pm 1.1$                   &    $2.3 \pm 1.3$                  &   $<$1.6 \\            
2003 WU$_{188}$  & 1342228922/...9040 & 226.1 & 2011-Sep-20 04:57 & 43.31 & 43.58 &  1.29 & $<$1.0                          &    $<$1.0                         & $<$1.1  \\      
\hline
1998 SN$_{165}$  & 1342212615/...2688 & 113.3 & 2011-Jan-15 00:39 & 37.71 & 37.95 &  1.46 & $9.5 \pm 1.2$                   &   $14.2 \pm 1.9$                  & $5.7 \pm 1.8$  \\
2001 QD$_{298}$  & 1342211949/...2033 & 188.5 & 2010-Dec-16 01:37 & 41.49 & 41.85 &  1.27 & $2.7 \pm 1.1$                   &   $4.1 \pm 1.3$                   & $<$1.3 \\
Altjira          & 1342190917/...1120 & 152.0 & 2010-Feb-23 00:32 & 45.54 & 45.58 &  1.25 & $4.5 \pm 1.4$                   &   $<$4.2                            & $<$2.3 \\

% 2003 MW$_{12}$   
Varda            & 1342213822/...3932 & 113.3 & 2011-Feb-08 06:52 & 47.62 & 47.99 &  1.11 & $23.1 \pm 1.7$                  &   $26.2 \pm 2.0$                  & $19.0 \pm 2.1$ \\
1999 RY$_{215}$  & 1342221751/...1778 & 188.5 & 2011-May-28 01:04 & 35.50 & 35.67 &  1.63 & $6.6 \pm 1.1$                   &   $5.7 \pm 1.6$                   & $<$2.4 \\
2002 AW$_{197}$  & 1342209471/...9654 & 113.3 & 2010-Nov-19 01:59 & 46.34 & 46.27 & 1.24  & $17.0 \pm 1.3$                  &   $20.2 \pm 1.7$                  & $15.3 \pm 1.5$ \\     
2005 UQ$_{513}$  & 1342212680/...2722 & 113.3 & 2011-Jan-15 20:26 & 48.65 & 48.80 & 1.16  & $5.3  \pm 1.5$                  &   $6.7 \pm 2.0$                   & $5.6 \pm 2.2$ \\ 
2004 PT$_{107}$  & 1342195396/...5462 & 113.3 & 2010-Apr-23 12:01 & 38.30 & 38.66 & 1.41  & $8.3  \pm 1.6$                  &   $8.6 \pm 2.2$                   & $7.9 \pm 2.8$  \\ 
2002 GH$_{32}$   & 1342212648/...2710 & 188.5 & 2011-Jan-15 11:35 & 43.29 & 43.64 & 1.22  & $<$1.1                          &   $<$1.5                          & $<$1.6  \\                 
2004 NT$_{33}$   & 1342219015/...9044 & 113.3 & 2011-Apr-19 07:34 & 38.33 & 38.69 & 1.42  & $17.3 \pm 1.7$                  &   $18.3 \pm 2.0$                  & $9.7 \pm 2.7$  \\                 
2004 XA$_{192}$  & 1342217343/...7399 &  75.7 & 2011-Mar-29 10:36 & 35.71 & 35.82 & 1.60  & $15.0 \pm 1.7$                  &   $14.2 \pm 2.2$                  & $8.2 \pm 5.9$  \\          
\hline
\end{tabular}
\label{table_obs}
\tablefoot{OBSIDs are the observation identifiers in the \emph{Herschel} Science Archive. Each target was observed four times
in visit 1 and four times in visit 2. The first OBSID of the consequtive four OBSIDs/visit are given. Duration is the total
duration of the two visits, mid-time is the mean UT time, $r$ is the mean heliocentric
distance, $\Delta$ is the mean {\it Herschel}-target distance, and $\alpha$ is
the mean Sun-target-{\it Herschel} phase angle (JPL Horizons Ephemeris System, \cite{Giorgini1996}).
Flux densities are color corrected and the 1$\sigma$ uncertainties include the absolute calibration
uncertainty. Upper limits are 1$\sigma$ noise levels of the final maps.
Targets below the horizontal line have i$>$$4.5\degr$.
}
\end{table*}

\subsection{\emph{Spitzer} observations}
\label{Sobs}
The Earth-trailing \emph{Spitzer Space Telescope} has a 0.85 m diameter
helium-cooled telescope. The cryogenic phase of the mission 
ended in 2009. During that phase, one of four science
instruments onboard, the
Multiband Imaging Photometer for Spitzer (MIPS; \cite{Rieke2004}), provided
useful photometry of TNOs at two bands: 24 and $70\ \mathrm{\mu m}$.
The latter is spectrally overlapping with the PACS $70\ \mathrm{\mu m}$ band
whereas the former can provide strong constraints on the temperature of the
warmest regions of TNOs. The telescope-limited spatial resolution is
$6\arcsec$ and $18\arcsec$ in the two bands, respectively. The nominal
absolute calibration, photometric methods, and color corrections are described
in \cite{Gordon2007}, \cite{Engelbracht2007} and \cite{Stansberry2007} (2007).
For TNOs we use larger calibration uncertainties of 3\% and 6\% at the 24
and $70\ \mathrm{\mu m}$ bands, respectively (\cite[2008]{Stansberry2008}).

\emph{Spitzer} 
observed about 100 TNOs and Centaurs and three-quarters of
them are also included in the ``TNOs are Cool'' \emph{Herschel} program. 
Many of the \emph{Spitzer} targets were observed multiple times within several days,
with the visits timed to allow subtraction of the background. A similar technique
has been applied also to the \emph{Herschel} observations
(Section~\ref{dataredux} / ``super-sky-subtraction'' method).
In this work and \cite{Vilenius2012} (2012) there are 
20 targets (out of 35 \emph{Herschel} targets analysed in these two works)
which have reanalyzed \emph{Spitzer}/MIPS data available
(Mueller et al., {\it in prep.}).
In addition, we have searched for all classical TNOs  observed with
\emph{Spitzer} but not with \emph{Herschel}: 1996 TS$_{66}$, 2001 CZ$_{31}$, 2001 QB$_{298}$,
2001 QC$_{298}$, 2002 GJ$_{32}$, 2002 VT$_{130}$, 2003 QR$_{91}$, and 2003 QY$_{90}$.
The dynamically hot CKBOs 1996 TS$_{66}$ and 2002 GJ$_{32}$ have been published in \cite{Brucker2009} (2009),
but their flux densities have been updated and reanalysed results of this work have changed their
size and albedo estimates (Table~\ref{collection}). 
An updated data reduction was recently done for 2001 QB$_{298}$ and 2002 VT$_{130}$ and
we use the results from \cite{Mommert2013} for these two targets.
Of the other targets
only 2001 QC$_{298}$ and 2003 QR$_{91}$ 
are finally used because all the other cases do
not have enough observations for a background removal or there was
a problem with the observation.
\emph{Spitzer} flux densities 
used in the current work
are given in Table~\ref{table_Spitzerobs}. For most of these targets
flux densities have been derived using multiple observations during an epoch
lasting one to eight days. Borasisi was observed in two epochs in 2004 and 2008.
The color corrections of CKBOs in our
sample are larger than in the case of the PACS instrument.
For MIPS the color corrections are 1\%-10\% of the flux density
at $24\ \mathrm{\mu m}$ and about 10\% at $70\ \mathrm{\mu m}$ obtained by a
method which uses the black body temperature which fits the 24:70 flux
ratio the best (\cite{Stansberry2007}, 2007).

\begin{table*}
 \centering
 \caption{\emph{Spitzer}/MIPS observations. Targets 2003 QR$_{91}$,  2001 QC$_{298}$,
 1996 TS$_{66}$ and 2002 GJ$_{32}$ were not observed by \emph{Herschel}. The latter
 two targets are from \cite{Brucker2009} (2009) and have been remodeled based on updated
 flux densities with significant changes in radiometric results. Flux densities of Teharonhiawako
 and 2001 QY$_{297}$ have been updated from those in \cite{Vilenius2012} (2012), and they have been
 reanalysed in this work.}
 \begin{tabular}{lrlccc|rc|rr}
  \hline
  Target        & PID   &  Mid-time of observation(s) & $r$   & $\Delta$ & $\alpha$ & \multicolumn{2}{c|}{MIPS $24\ \mathrm{\mu m}$ band} & \multicolumn{2}{c}{MIPS $70\ \mathrm{\mu m}$ band} \\
                &       &                             & (AU)  & (AU)     & (\degr)  & Dur. (min)    & F$_{24}$ (mJy) & Dur. (min)        & F$_{70}$ (mJy)     \\
  \hline
2001 QS$_{322}$ & 3542  &  2005-Dec-03 13:12  & 42.32 & 41.87    & 1.23 & 467.2       & $<0.015$  & 308.5      & $<1.1$  \\
Borasisi        & 3229  &  2004-Dec-02 00:29  & 41.16 & 41.16    & 1.41 & 99.1        & $<0.030$  & 218.2      & $3.6 \pm 0.9$ \\
                & 50024 &  2008-Jul-29 16:48 & 41.41 & 40.97    & 1.29 & 170.45      & $0.057 \pm 0.007$  & 369.3      & $1.3 \pm 0.7$ \\                
2001 QY$_{297}$ & 50024 &  2008-Nov-25 02:40 & 43.09 & 42.73    & 1.28 & 170.45      & $0.016 \pm 0.006$  & 239.17     & $2.0 \pm 0.9$ \\
2001 QT$_{322}$ & 3542  &  2004-Dec-26 23:46 & 36.92 & 36.95    & 1.56 & 406.6       & $<0.037$           & 406.6      & $<1.5$         \\
2003 QA$_{91}$  & 50024 &  2008-Dec-28 16:34 & 44.91 & 44.87    & 1.29 & 431.8       & $0.079 \pm 0.006$  & 639.2      & $3.8 \pm 0.4$ \\
Teharonhiawako  & 3229  &  2004-Nov-09 20:15 & 45.00 & 44.72    & 1.25 & 153.27      & $0.027 \pm 0.010$  & 179.03     & $1.6 \pm 0.6$ \\
2003 QR$_{91}$  & 50024 &  2008-Nov-24 13:26 & 39.12 & 38.70    & 1.37 & 340.9       & $0.107 \pm 0.007$  & 1074.2     & $4.8 \pm 0.6$ \\
\hline
1998 SN$_{165}$ & 55    &  2004-Dec-05 08:10 & 37.97 & 37.54    & 1.39 & \multicolumn{2}{c|}{no observations}                  & 37.3      & $<13.9$ \\
2001 QD$_{298}$ & 3542  &  2004-Nov-05 13:41 & 41.19 & 40.91    & 1.36 & 283.3       & $<0.059$           & 283.3     & $<1.5$   \\
1996 TS$_{66}$  & 3542  &  2005-Jan-29 06:34 & 38.53 & 38.21    & 1.42 & 114.96      & $0.104 \pm 0.009$  & 268.54    & $2.3 \pm 0.8$ \\
2002 GJ$_{32}$  & 3542  &  2006-Feb-19 07:31 & 43.16 & 43.16    & 1.33 & 214.06      & $0.024 \pm 0.006$  & 132.87    & $4.2 \pm 0.9$ \\
2002 AW$_{197}$ & 55    &  2004-Apr-12 16:34 & 47.13 & 46.70    & 1.10 & 56.7        & $0.143 \pm 0.027$ & 56.7      & $13.7 \pm 1.9$ \\
2001 QC$_{298}$ & 50024 &  2008-Jul-29 20:53 & 40.62 & 40.31    & 1.38 & 170.45      & $0.158 \pm 0.010$ & 369.25    & $5.8 \pm 0.7$ \\
\hline
 \end{tabular}
\label{table_Spitzerobs}
\tablefoot{PID is the \emph{Spitzer} program identifier. 
Observing geometry (heliocentric distance $r$,
\emph{Spitzer}-target distance $\Delta$ and Sun-target-\emph{Spitzer} phase angle $\alpha$) is averaged
over the individual observations. The ``Dur.'' column gives 
the total observing time of several visits. The durations of observing epochs were 4-8 days,
except for 1998 SN$_{165}$ and 2002 AW$_{197}$, which had only one observation.
The effective monochromatic wavelengths of the two MIPS bands we use are $23.68\ \mathrm{\mu m}$
and $71.42\ \mathrm{\mu m}$. Targets below the horizontal line have $i>4.5\degr$. 
\\
\\
{\bf References.}
In-band fluxes from Mueller et al., ({\it in prep}.). Flux densities presented in this table
have been color corrected.}
\end{table*}

\subsection{Optical data}
\label{auxobs}
We use the V-band absolute magnitudes ($H_{\mathrm{V}}$ as given in Table~\ref{table_overview_groundobs})
as input in the modeling (Section \ref{model}).
The quantity and quality of published $H_{\mathrm{V}}$s or individual V-band or R-band
observations vary significantly for our sample. 
Some of our targets have been observed in the Sloan Digital Sky Survey
and to convert from their r and g bands to V-band we use the
transformation\footnote{\url{http://www.sdss.org/dr5/algorithms/sdssUBVRITransform.html}, accessed February 2013.}
\begin{equation}
V = g-0.5784 \left( g-r \right) -0.0038.
\label{SLOANtransform}
\end{equation}
The estimated uncertainty of this transformation is 0.02 mag.

To take into account brightening at small
phase angles we use the linear method commonly used for distant Solar System objects:
\begin{equation}
 H_{\mathrm{V}}=V-5 \log \left( r \Delta \right)-\beta_{\mathrm{V}} \alpha,
\label{Hmag}
\end{equation}
where $r$ is the heliocentric distance, $\Delta$ the observer-target
distance, $\beta_{\mathrm{V}}$ the linear phase coefficient in V-band, and $\alpha$ the Sun-target-observer
phase angle. Often the linear phase coefficient cannot be deduced in
a reliable way from the few data points available and in those cases we use as default the average
values $\beta_{\mathrm{V}}$\,=\,$0.112 \pm 0.022$ or $\beta_{\mathrm{R}}$\,=\,$0.119 \pm 0.029$
(\cite[2008]{Belskaya2008}). Many published $H_{\mathrm{V}}$ values are also based on an assumed
phase coefficient.
We prefer to use mainly published photometric quality observations due to their careful calibration
and good repeatability. For each target we try to determine $H_\mathrm{V}$ and $\beta$ by making a fit to
the combined V-data collected from literature. We have determined new linear phase coefficients of
Borasisi: $\beta_{\mathrm{V}}$\,=\,$0.176 \pm 0.073$, 1998 SN$_{165}$:
$\beta_{\mathrm{V}}$\,=\,$0.132 \pm 0.063$ and 2001 QC$_{298}$: $\beta_{\mathrm{V}}$\,=\,$1.01 \pm 0.29$.

When no other sources are available, or the high-quality data is based on one or two data points,
we also take into account data from the
Minor Planet Center (MPC). These observations are often more numerous, or only available at, the R-band.
We check if consistency and phase angle coverage of MPC data allow to fit the slope (i.e. $\beta$)
in a reliable way, otherwise the fit is done using the default phase coefficient.
Unless available for a specific target (Table~\ref{table_overview}), we use the average (V-R)
color index for CKBOs, which has been determined separately for cold
and hot classicals\footnote{Note that \cite{Vilenius2012} (2012) used one average
in their analysis of \emph{Herschel} data on classical TNOs: V-R=$0.59 \pm 0.15$ based on an earlier
version of the MBOSS data base.}.
The average of 49 cold CKBOs is V-R=0.63\,$\pm$\,0.09 and of 43 hot CKBOs V-R=$0.51 \pm 0.14$ (\cite[2012]{Hainaut2012}).
The MPC is mainly used for astrometry and can differ significantly from
well-calibrated photometry. Comparisons by \cite{Romanishin2005} (2005) and \cite{Benecchi2011} (2011)
indicate an offset of $\sim$\,$0.3$ mag (MPC having brighter magnitudes) with a scatter of $\sim$\,$0.3$ mag.
We have assigned an uncertainty of 0.6 mag to MPC data points.
The absolute magnitudes and their error bars used as input in our analysis
(Table~\ref{table_overview_groundobs})
take into account additional uncertainties from
known or assumed light curve variability in $H_{\mathrm{V}}$ as explained in \cite{Vilenius2012} (2012).

\begin{table*}
\centering
\caption{Optical auxiliary data based on a linear phase curve fit using V-band data points from the literature or data bases.}
\begin{tabular}{lccrccccc}
\hline
\hline
Target                           &   & $H_{\mathrm{V}}$   &  N & Phase coefficient            & L.c. $\Delta m_\mathrm{R}$        & L.c. period    & $H_{\mathrm{V}}$  & Comment \\
                                 &   & ref.               &    & (mag/\degr)                  & (mag)                             & ($\mathrm{h}$) & (mag)             &    \\
\hline
(2001 QS$_{322}$)                &   & (x)                &  4 & (default)                    & \ldots                            & \ldots         & $6.91 \pm 0.68$   & Default V-R \\
66652 Borasisi (1999 RZ$_{253}$) & B & (e,l,f,y)          &  7 & $0.176 \pm 0.073$            & $0.08\pm 0.02$\tablefootmark{z}   & $6.4\pm 1.0$\tablefootmark{z} & $6.121 \pm 0.070$ & New $\beta$ fit \\
(2003 GH$_{55}$)                 &   & (c)                &  3 & (default)                    & \ldots                            & \ldots         & $6.43 \pm 0.12$   &  \\
135182 (2001 QT$_{322}$)         &   & (h,x)              &  5 & (default)                    & \ldots                            & \ldots         & $7.29 \pm 0.67$   & V-R from (h) \\
(2003 QA$_{91}$)                 & B & (x)                & 13 & (default)                    & \ldots                            & \ldots         & $5.76 \pm 0.63$   & Default V-R \\
(2003 QR$_{91}$)                 & B & (x)                &  8 & (default)                    & \ldots                            & \ldots         & $6.55 \pm 0.56$   & Default V-R \\
(2003 WU$_{188}$)                & B & (x)                &  8 & (default)                    & \ldots                            & \ldots         & $5.96 \pm 0.64$   & Default V-R \\
\hline
35671 (1998 SN$_{165}$)          &   & (j,k,l,y,b2)       & 20 & $0.146 \pm 0.063$            & $0.16 \pm 0.01$\tablefootmark{a2}  & 8.84\tablefootmark{a2} & $5.707 \pm 0.085$ & New $\beta$ fit \\
(2001 QD$_{298}$)                &   & (m)                &  1 & (default)                    & \ldots                            & \ldots         & $6.71 \pm 0.17$ & \\ 
174567 Varda (2003 MW$_{12}$)    & B & (c)                &  6 & (default)                    & $0.06 \pm 0.01$\tablefootmark{c2} & 5.9\tablefootmark{c2} & $3.61 \pm 0.05$ & \\  
86177 (1999 RY$_{215}$)          &   & (c)                &  1 & (default)                    & $<\,0.1$\tablefootmark{v}         & \ldots         & $7.235 \pm 0.093$ & \\  
55565 (2002 AW$_{197}$)          &   & (s)                & \multicolumn{2}{l}{(phase curve study)} & $0.08 \pm 0.07$\tablefootmark{d2} & $8.86 \pm 0.01$\tablefootmark{d2} & $3.568 \pm 0.046$ & \\  
202421 (2005 UQ$_{513}$)         &   & (v)                & 10 & (default)                    & $0.06 \pm 0.02$\tablefootmark{e2} & 7.03\tablefootmark{e2} & $3.87 \pm 0.14$ & Default V-R \\  
(2004 PT$_{107}$)                &   & (v)                & 24 & (default)                    & $0.05 \pm 0.1$\tablefootmark{v}   & $\sim$20\tablefootmark{v} & $6.33 \pm 0.11$ & Default V-R \\  
(2002 GH$_{32}$)                 &   & (m,w)              &  2 & (default)                    & \ldots                            & \ldots & $6.58 \pm 0.28$ & V-R from (w)\\  
2001 QC$_{298}$                  & B & (e,g,v)            &  3 & $1.01 \pm 0.29$              & $0.4 \pm 0.1$\tablefootmark{v}    & $\sim$12\tablefootmark{v} & $6.26 \pm 0.32$ & Default V-R \\
(2004 NT$_{33}$)                 &   & (c)                &  6 & (default)                    & $0.04 \pm 0.01$\tablefootmark{e2} & 7.87\tablefootmark{e2}    & $4.74 \pm 0.11$ & \\  
230965 (2004 XA$_{192}$)         &   & (x)                & 17 & (default)                    & $0.07 \pm 0.02$\tablefootmark{e2} & 7.88\tablefootmark{e2}    & $4.42 \pm 0.63$ & Default V-R \\
\hline
\end{tabular}
\label{table_overview_groundobs}
\tablefoot{B denotes a known binary system (\cite{Noll2008}, \cite{Noll2009}, \cite{Benecchi2013}), $N$ is the total number of individual V or
R-band data points used, the phase coefficient is explained in the text and Equation~(\ref{Hmag}), 
$H_{\mathrm{V}}$ are the absolute
V-band magnitudes with uncertainties taking into account
lightcurve (L.c.) amplitude $\Delta m_R$.
Targets below the horizontal line have inclinations $>$4.5$\degr$.
\\
\\
{\bf References.} (c)-(w) given below Table~\ref{table_overview}.
\tablefoottext{x}{R-band data from IAU Minor Planet Center, \url{http://www.minorplanetcenter.net/db_search/}, accessed July 2012.}
\tablefoottext{y}{\cite{McBride2003}.}
\tablefoottext{z}{\cite{Kern2006}.}
\tablefoottext{a2}{\cite{Lacerda2006}.}
\tablefoottext{b2}{From \cite{Ofek2012} using Equation~(\ref{SLOANtransform}).}
\tablefoottext{c2}{\cite{Thirouin2010} (2010).}
\tablefoottext{d2}{\cite{Ortiz2006}.}
\tablefoottext{e2}{\cite{Thirouin2012} (2012).}
}
\end{table*}

\section{Analysis}
\subsection{Thermal modeling}
\label{model}
We aim to solve for size (effective diameter $D$ assuming spherical shapes),
geometric albedo $p_\mathrm{V}$ and beaming factor $\eta$
by fitting the 
two or more 
thermal infrared data points as well as the optical
$H_{\mathrm{V}}$ data in the pair of equations
\begin{equation}
F(\lambda) = \frac{\epsilon\left(\lambda\right)}{\Delta^2}
\int_{S} B \left( \lambda, T\left( S,\eta \right)\right) \;d\textbf{S} \cdot \textbf{u}
\label{model_emission}
\end{equation}
\begin{equation}
H_\mathrm{V} = m_\mathrm{\sun}+5 \log \left(\sqrt{\pi} a \right)-\frac{5}{2} \log \left(p_\mathrm{V} S_\mathrm{proj} \right),
\label{opt_constr}
\end{equation}
where $F$ is the flux density, $\lambda$ the wavelength, $\epsilon$ the emissivity,
$\Delta$ the observer-target distance,
$B(\lambda, T)$ Planck's radiation law for black bodies, $T \left( S,\eta \right)$
the temperature distribution on the surface $S$ adjusted by the beaming factor $\eta$, $\textbf{u}$ the
unit directional vector towards the observer from the surface element $d\textbf{S}$,
$m_\mathrm{\sun}$ the apparent magnitude of the Sun, $a$ the distance of one astronomical unit
and $S_\mathrm{proj}$ the area of the target projected towards the observer.
To model the temperature distribution on the surface of an airless, spherical TNO
we use the Near-Earth Asteroid Thermal Model NEATM (\cite{Harris1998}). For a description
of our NEATM implementation for TNOs we refer to \cite{Mommert2012} (2012).
The temperature distribution across an
object differs from the temperature distribution which a smooth object
in instantaneous equilibrium with insolation would have. This adjustment
is done by the beaming factor $\eta$ which scales the temperature as
$T\propto \eta^{-0.25}$. In addition to the quantities explicitly
used in NEATM (solar flux, albedo, heliocentric distance, emissivity) the
temperature distribution is affected by other effects combined in $\eta$:
thermal inertia, surface roughness and the rotation state of the object.
Statistically, without detailed information about the spin-axis orientation and period,
large $\eta$ indicates high thermal inertia, and $\eta$$<$1 indicates a rough surface.
Thermal properties of TNOs have been analysed in detail by \cite{Lellouch2013} (2013).

Emissivity is assumed to
be constant 
$\epsilon\left(\lambda\right)=0.9$ 
as discussed
in \cite{Vilenius2012} (2012). This assumption is often used for small
Solar System bodies. A recent \emph{Herschel} study using both PACS and
SPIRE instruments (70, 100, 160, 250, 350 and $500\ \mathrm{\mu m}$ photometric bands) shows that
in a sample of nine TNOs/Centaurs most targets 
show significant indications of an emissivity
decrease, but only at wavelenghts above $250\ \mathrm{\mu m}$, except for
one active Centaur (\cite{Fornasier2013} 2013). Thus, we assume that
emissivity of CKBOs is constant at MIPS and PACS wavelengths.

The free parameters $p_V$, $D=\sqrt{\frac{4 S_\mathrm{proj}}{\pi}}$ and $\eta$ are
fitted in a weighted least-squares sense by minimizing
\begin{equation}
 \chi_{\nu}^2=\frac{1}{\nu} \sum_{i=1}^{N} \frac{\left[ F \left( \lambda_i \right) -
 F_{\mathrm{model}} \left( \lambda_i \right)\right]^2}{\sigma_i^2},
\label{chi2}
\end{equation}
where $\chi_{\nu}^2$ is called the ``reduced $\chi^2$'', $\nu$ is the number of degrees of freedom,
$N$ the number of data points,
$F \left( \lambda_i \right)$ the observed flux density at wavelength $\lambda_i$, or $H_\mathrm{V}$
transformed to flux density scale, with uncertainty
$\sigma_i$, and $F_{\mathrm{model}}$ is the 
calculated thermal emission or optical brightness from Eqs.~\ref{model_emission} and \ref{opt_constr}.
The number of degrees of freedom is $N$-3 
when $H_\mathrm{V}$ is counted as one data point.
If the fit fails or gives an unphysical $\eta$ then a fixed-$\eta$ fit is made instead
(see Section~\ref{resultsection}) and the number of degrees of freedom is $N$-2.

The error estimates of the
fitted parameters are determined by a Monte Carlo method (\cite[2011]{Mueller2011})
using a set of 1000 randomized input flux densities and absolute visual magnitudes
for each target, as well as beaming factors for fixed-$\eta$ cases. Our implementation
of the technique is shown in \cite{Mommert2012} (2012). In cases of poor fit,
i.e. reduced-$\chi^2$ significantly greater than one, the error bars are first
rescaled so that the Monte Carlo method would not underestimate the uncertainties
of the fitted parameters.
This is discussed in \cite{SantosSanz2012} (2012, Appendix B.1.).
The assumption that the targets are spherical may slightly overestimate diameters,
since most TNOs are known to be MacLaurin spheroids (\cite{Duffard2009} 2009, \cite{Thirouin2010} 2010).
NEATM model accuracy at small phase angles is about 5\% in the diameter estimates and
10\% in the geometric albedo (e.g. \cite{Harris2006}).

\subsubsection{Treatment of upper limits}
\label{upplims}
Tables~\ref{table_obs} and \ref{table_Spitzerobs} list several data points
where only an upper limit for flux density is given.
As mentioned in Section~\ref{dataredux} the observed flux densities of our
sample were often lower than predictions by a factor of two or more. In the
planning we aimed at SNR=2-4 for the faintest targets (Section~\ref{Hobs}).
If a target has at least one SNR$>$1 data point we can
assume that the flux densities are not far below the SNR=1 detection limit in the other,
non-detected, bands. 
Such upper limits we replace by a distribution of possible flux densities.
We assign them
values,
using a Monte Carlo technique, from a one-sided Gaussian distribution with the map noise
(upper limits in Tables \ref{table_obs} and \ref{table_Spitzerobs}) as the standard deviation.
We calculate the optimum solution in the sense of Eq.~\ref{chi2} and repeat this 500 times.
The adopted $D$, $p_\mathrm{V}$ and $\eta$ are the medians of all the obtained values of the
fitted parameters, respectively.

It should be noted that 
both the treatment of upper limit bands as well as non-detected targets (discussed below)
is done in
a different way in this work than in 
previous works who treated
upper limits as data points with zero flux density: 0$\pm 1\sigma$.
We have remodeled the CKBO sample of \cite{Vilenius2012} (2012)
using our new 
convention and find changes in size larger than $\sim$10\% for a few   
targets 
(see Section~\ref{resultscomp}).

For targets which are non-detections in all bands we give
upper limits for diameters and lower limits for geometric albedos. We calculate them by making
a fixed-$\eta$ fit to the most constraining
upper limit and assign a zero flux density in that band, 
which is the $70\,\mu m$ band in all the three cases
(2002 GV$_{31}$, 2003 WU$_{188}$, 2002 GH$_{32}$), 
using a $2\sigma$ uncertainty. 
The reason to choose $2\sigma$ instead of $1\sigma$ 
for non-detections
is explained in the following.
At the limit of detection SNR=1 and we have a flux density of
$F$$=$s$\pm$s, where s is the 1$\sigma$ Gaussian noise level of the map
determined by doing photometry on 200 artificial
sources randomly implanted near the target. Thus, the probability
that the ``true'' flux density of the target is more than 1$\sigma$
above the nominal value s (i.e. $F$$>$2s) is 16\%. On the other hand, if the SNR=1 observation
is interpreted as an upper limit
a similar probability for the flux density to exceed $F$$>$2s should occur. This
requires that upper limits, which have been assigned zero flux for non-detections,
are treated as 0$\pm 2\sigma$ in order to avoid this discontinuity at SNR=1.

\subsection{Results of model fits}
\label{resultsection}
The results of model fits using the NEATM (see Section~\ref{model}) are given in
Table~\ref{table_results}.
For binary systems the diameters are to be interpreted as
area-equivalent diameters because our observations did not spatially resolve
separate components. The prefered solutions, based on the combination of
\emph{Herschel}/PACS and \emph{Spitzer}/MIPS data when available, are shown in
Fig.~\ref{fits_lot1}. Although size estimates can be done using one instrument alone,
the combination of both instruments samples the thermal peak
and the short-wavelength side of the SED by extending the wavelength coverage and
number of data points. When possible, we solve for three parameters:
radiometric (system) diameter, geometric albedo
and beaming factor. If data consistency does not allow a three-parameter solution we
fit for diameter and albedo.
This type of ``fixed-$\eta$'' solution is chosen if a floating-$\eta$ solution (i.e. $\eta$ as one
of the parameters to be fitted) gives an ``unphysical'' beaming factor ($\eta$$\lesssim$0.6 or
$\eta$$>$2.6). An often used value for the fixed-$\eta$ is 1.20$\pm$0.35
(\cite[2008]{Stansberry2008}) and it was used also in previous works based on
\emph{Herschel} data (\cite[2012]{SantosSanz2012}, \cite[2012]{Mommert2012},
\cite[2012]{Vilenius2012}). 
A three-parameter fit may give a solution which has very large error bars
such that the uncertainty in $\eta$ would cover its whole physical range. In such cases
we have adopted the fitted value of $\eta$ as an ``adjusted fixed-$\eta$'' value and run
the fit again keeping $\eta$ constant. In these  cases we assign an error bar of $\pm 0.35$
to the ``adjusted fixed-$\eta$'' value to be consistent with estimates produced with
the default fixed eta of 1.20$\pm$0.35. The type of solution is indicated in
Table~\ref{table_results}.

Since 
many of our targets 
have data only from 
PACS we show also
the PACS-only solutions in Table~\ref{table_results} 
for all targets which have been detected in at least one PACS band.
In many cases the data from PACS and the combined data set are consistent with each other
and the difference is small. An exception is 2001 QS$_{322}$. For this target
the different solutions 
are due
to the effect of the 24 $\mu m$ MIPS upper limit.

\begin{table*}
\centering
\caption{Solutions of radiometric modeling. The prefered solution (target name and instruments in boldface) is the one with 
data from two instruments, when available (see also Section~\ref{resultsection}).}
\begin{tabular}{lllcccccc}
\hline
Target                           &   & Instruments & No. of & $D$               & $p_\mathrm{V}$ \tablefootmark{(a)}      & $\eta$ & Solution & Comment  \\
                                 &   &             & bands  & (km)              &                                         &        & type     &          \\
\hline
\hline
(2001 QS$_{322}$)                &   & PACS        & 3       & $253_{-29}^{+87}$ & $0.048_{-0.030}^{+0.587}$ & $1.20 \pm 0.35$ & fixed $\eta$ & default $\eta$ \\  
\noalign{\smallskip}

\textbf{(2001 QS$_{\textbf{322}}$)} &   & \textbf{PACS, MIPS}  & 5       & $186_{-24}^{+99}$ & $0.095_{-0.060}^{+0.531}$ & $1.20 \pm 0.35$ & fixed $\eta$ & default $\eta$ \\

                                 &   &             &         &                   &                           &         &  & \\

\textbf{66652 Borasisi (1999 RZ$_{\textbf{253}}$)} & B & \textbf{PACS, MIPS}  & 5       & $163_{-66}^{+32}$ & $0.236_{-0.077}^{+0.438}$ & $0.77_{-0.47}^{+0.19}$ & floating $\eta$ & \\ 

                                 &   &             &         &                   &                           &         &  & \\

\textbf{(2003 GH$_{\textbf{55}}$)}                 &   & \textbf{PACS}        & 3       & $178_{-56}^{+21}$ & $0.150_{-0.031}^{+0.182}$ & $1.20 \pm 0.35$ & fixed $\eta$ & default $\eta$  \\

                                 &   &             &         &                   &                           &         &   \\
135182 (2001 QT$_{322}$)         &   & PACS        & 3       & $173_{-55}^{+25}$ & $0.071_{-0.044}^{+0.091}$ & $1.20 \pm 0.35$ & fixed $\eta$ & default $\eta$  \\ 
\noalign{\smallskip}

\textbf{135182 (2001 QT$_{\textbf{322}}$)}  &   & \textbf{PACS, MIPS}  & 5       & $159_{-47}^{+30}$ & $0.085_{-0.052}^{+0.424}$ & $1.20 \pm 0.35$ & fixed $\eta$ & default $\eta$ \\

                                 &   &             &         &                   &                           &         &  & \\
(2003 QA$_{91}$)                 & B & PACS        & 3       & $233_{-56}^{+40}$ & $0.162_{-0.094}^{+0.162}$ & $1.20 \pm 0.35$ & fixed $\eta$ & default $\eta$  \\ 
\noalign{\smallskip}

\textbf{(2003 QA$_{\textbf{91}}$)}        &   & \textbf{PACS, MIPS}  & 5       & $260_{-36}^{+30}$ & $0.130_{-0.075}^{+0.119}$ & $0.83_{-0.15}^{+0.10}$ & floating $\eta$ & \\

                                 &   &             &         &                   &                           &         & &  \\

\textbf{(2003 QR$_{\textbf{91}}$)}                 & B & \textbf{MIPS}        & 2       & $280_{-30}^{+27}$ & $0.054_{-0.028}^{+0.035}$  & $1.20_{-0.12}^{+0.10}$  & floating $\eta$ & \\

                                 &   &             &         &                   &                      &         & &  \\
\textbf{(2003 WU$_{\textbf{188}}$)}                & B & \textbf{PACS}        & 3       & $<$220 & $>$0.15 & $1.20 \pm 0.35$ & fixed $\eta$ & default $\eta$  \\

\noalign{\smallskip}
\hline\noalign{\smallskip}
35671 (1998 SN$_{165}$)          &   & PACS        & 3       & $392_{-52}^{+43}$ & $0.060_{-0.012}^{+0.020}$ & $1.22 \pm 0.35$ & fixed $\eta$ & adjusted $\eta$ \\ 
\noalign{\smallskip}

\textbf{35671 (1998 SN$_{\textbf{165}}$)} &   & \textbf{PACS, MIPS}  & 4       & $393_{-48}^{+49}$      & $0.060_{-0.013}^{+0.019}$ & $1.23 \pm 0.35$ & fixed $\eta$ & adjusted $\eta$ \\

                                 &   &             &         &                   &                        &         &  & \\
(2001 QD$_{298}$)                &   & PACS        & 3       & $237_{-53}^{+25}$ & $0.065_{-0.013}^{+0.039}$ & $1.20 \pm 0.35$ & fixed $\eta$ & default $\eta$  \\

\noalign{\smallskip}
\textbf{(2001 QD$_{\textbf{298}}$)}                                 &   & \textbf{PACS, MIPS}  & 5       & $233_{-63}^{+27}$ & $0.067_{-0.014}^{+0.062}$ & $1.26 \pm 0.35$ & fixed $\eta$ & adjusted $\eta$ \\ 
                                 &   &             &         &                   &                      &         &  & \\
\textbf{174567 Varda (2003 MW$_{\textbf{12}}$)}    & B & \textbf{PACS}        & 3       & $792_{-84}^{+91}$ & $0.102_{-0.020}^{+0.024}$ & $0.84_{-0.22}^{+0.28}$ & floating $\eta$ & \\ 

                                 &   &             &         &                   &                      &         &  & \\

\textbf{86177 (1999 RY$_{215}$)}          &   & \textbf{PACS}        & 3       & $263_{-37}^{+29}$ & $0.0325_{-0.0065}^{+0.0122}$   & $1.20 \pm 0.35$ & fixed $\eta$ & default $\eta$ \\ 

                                 &   &             &         &                   &                      &         &  & \\
55565 (2002 AW$_{197}$)          &   & PACS        & 3       & $714_{-74}^{+76}$ & $0.130_{-0.023}^{+0.031}$      & $1.04_{-0.27}^{+0.31}$    & floating $\eta$ & \\ 
 \noalign{\smallskip}

\textbf{55565 (2002 AW$_{\textbf{197}}$)}                                  &   & \textbf{PACS, MIPS}  & 5       & $768_{-38}^{+39}$ \tablefootmark{(b)} & $0.112_{-0.011}^{+0.012}$      & $1.29_{-0.10}^{+0.13}$  & floating $\eta$ & \\ 
                                 
                                 &   &             &         &                   &                      &         & &  \\
\textbf{202421 (2005 UQ$_{\textbf{513}}$)}         &   & \textbf{PACS}        & 3       & $498_{-75}^{+63}$ & $0.202_{-0.049}^{+0.084}$    & $1.27 \pm 0.35$ & fixed $\eta$ & adjusted $\eta$ \\

                                 &   &             &         &                   &                      &         &  & \\

\textbf{(2004 PT$_{\textbf{107}}$)}                &   & \textbf{PACS}        & 3       & $400_{-51}^{+45}$ & $0.0325_{-0.0066}^{+0.0111}$    & $1.53 \pm 0.35$ & fixed $\eta$ & adjusted $\eta$ \\

                                 &   &             &         &                   &                      &         & &  \\

\textbf{(2002 GH$_{\textbf{32}}$)}                 &   & \textbf{PACS}        & 3       & $<$\,230          & $>$\,0.075           & $1.20 \pm 0.35$ & fixed $\eta$ & default $\eta$ \\

                                 &   &             &         &                   &                      &         &  & \\

\textbf{2001 QC$_{\textbf{298}}$}                  & B  & \textbf{MIPS}       & 2       & $303_{-32}^{+29}$ & $0.063_{-0.018}^{+0.029}$ & $0.983_{-0.097}^{+0.085}$ & floating $\eta$ & \\

                                 &   &             &         &                   &                      &         &  & \\
\textbf{(2004 NT$_{\textbf{33}}$)}                 &   & \textbf{PACS}        & 3       & $423_{-80}^{+87}$ & $0.125_{-0.039}^{+0.069}$    & $0.69_{-0.32}^{+0.46}$ & floating $\eta$ & \\ 
                                 &   &             &         &                   &                      &         &  & \\
\textbf{230965 (2004 XA$_{\textbf{192}}$)}         &   & \textbf{PACS}        & 3       & $339_{-95}^{+120}$ & $0.26_{-0.15}^{+0.34}$      & $0.62_{-0.49}^{+0.79}$ & floating $\eta$ & \\ 
\hline
\end{tabular}
\label{table_results}
\tablefoot{'B' indicates a known binary system and the diameter given is the area-equivalent system diameter.
\tablefoottext{a}{Lower uncertainty limited by the uncertainty of $H_{\mathrm{V}}$ for 2001 QS$_{322}$ (both solutions),
2003 QA$_{91}$ (both solutions), 2003 QR$_{91}$, 2001 QT$_{322}$ (both solutions),
2001 QC$_{298}$, and 2004 XA$_{192}$.} 
\tablefoottext{b}{Lower uncertainty limited by the diameter uncertainty of 5\% of the NEATM model.}
}
\end{table*}

\begin{figure*}
   \includegraphics[width=17cm]{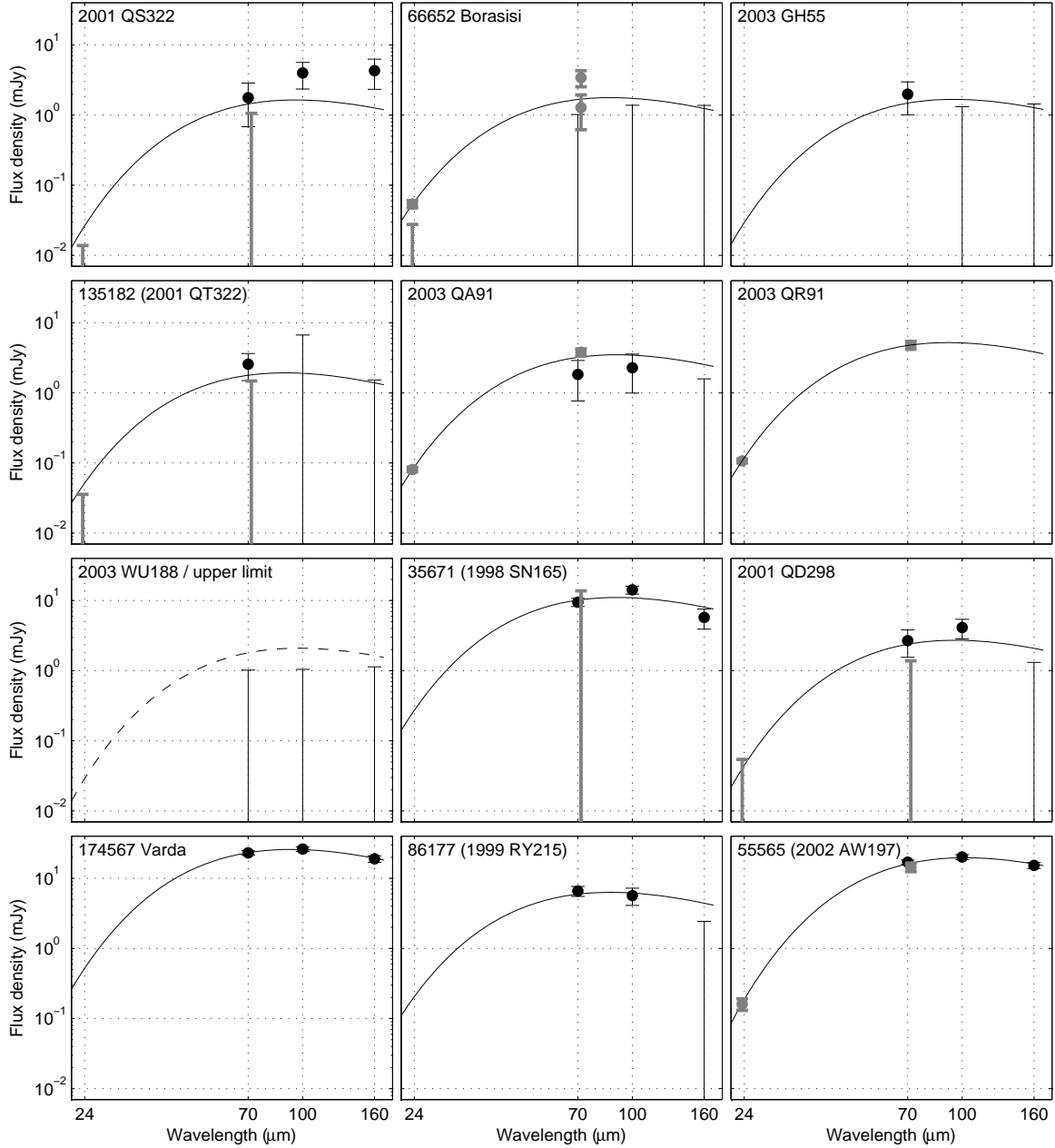}   
   \caption{SEDs calculated from the model solutions (Table~\ref{table_results}). The black data points are from PACS
(70, 100 and $160\ \mathrm{\mu m}$) and the gray points are from MIPS
(23.68 and $71.42\ \mathrm{\mu m}$) normalized to the observing geometry of PACS.
Error bars without a data point indicate 1$\sigma$ upper limits. An upper-limit solution based on a non-detection
is marked with a dashed line (see text). Target 2003 QR$_{91}$ was not observed by PACS.
}
   \label{fits_lot1}
\end{figure*}
\setcounter{figure}{0} 
\begin{figure*}
   \includegraphics[width=17cm]{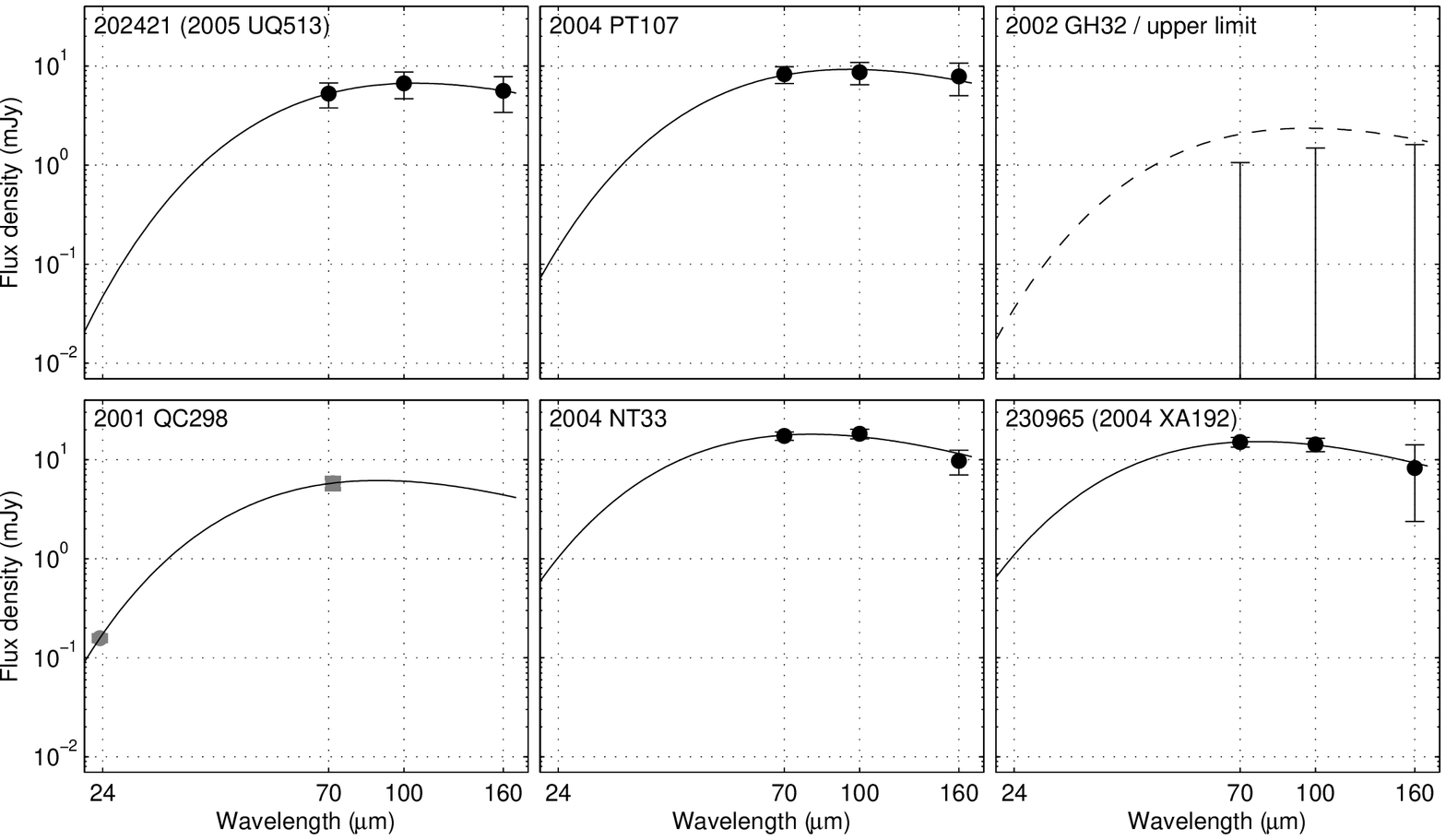}   
   \caption{continued. 2002 GH$_{32}$ has an upper limit solution (see text).}
   \vspace{15mm}
   \label{fits_lot2}
\end{figure*}

\subsection{Comparison with earlier results}
\label{resultscomp}
Of the 
18 targets in our sample only 2001 QD$_{298}$ and 2002 AW$_{197}$ have
earlier published diameter/albedo solutions and additionally
2001 QS$_{322}$ and 2001 QT$_{322}$ have upper size limits 
in the literature.
For 2001 QD$_{298}$ the \emph{Spitzer}/MIPS based result, with 
different MIPS flux densities and $H_\mathrm{V}$ than used in this work 
(Table~\ref{table_Spitzerobs}), 
was $D$$=$$150_{-40}^{+50}$ km, $p_\mathrm{V}$$=$$0.18_{-0.08}^{+0.17}$,
$\eta$$=$$0.79_{-0.26}^{+0.28}$ (\cite[2009]{Brucker2009}).
Our new diameter ($233_{-63}^{+27}$ km) 
is larger and geometric albedo ($0.067_{-0.014}^{+0.062}$) is lower than the previous
estimate. 

The first size measurement of 2002 AW$_{197}$ was done with the Max Planck Millimeter Bolometer
at the IRAM 30 m telescope. The result of \cite{Margot2002} (2002) was $D$$=$$886_{-131}^{+115}$ km
and $p_\mathrm{R}$$=$$0.101_{-0.022}^{+0.038}$. \emph{Spitzer} measurements 
gave a smaller size $D$$=$$740\pm$$100$ km and $p_\mathrm{V}$$=$$0.12_{-0.03}^{+0.04}$
(\cite[2009]{Brucker2009}).
Our new result is close to this 
and has significantly smaller error bars ($D$$=$$768_{-38}^{+39}$ km, $p_\mathrm{V}$$=$$0.112_{-0.011}^{+0.012}$).

The previous limits of
2001 QS$_{322}$ were $D$$<$200 km and $p_\mathrm{V}$$>$$0.15$ (\cite[2009]{Brucker2009}). 
While the diameter limit is
compatible with the new size estimate ($186_{-24}^{+99}$ km) the new geometric
albedo is lower ($0.095_{-0.060}^{+0.531}$) due to PACS data points and
updated $H_\mathrm{V}$. Also the MIPS data has been reanalysed and has changed for
this target. Similarly, the geometric albedo estimate of 2001 QT$_{322}$ is now
$0.085_{-0.052}^{+0.424}$ which is lower than the previous lower limit
of 0.21 (\cite[2009]{Brucker2009}). 
We use a different absolute visual magnitude $H_\mathrm{V}$$=$$7.29 \pm 0.67$, whereas \cite{Brucker2009}
(2009) used $6.4 \pm 0.5$ for 2001 QT$_{322}$.

For binary targets it is possible to estimate a size range based on
the assumptions of spherical shapes and equal albedos of the
primary and secondary components. Assuming a bulk density range of
0.5-2.0 g cm$^{-3}$ and using the system mass and brightness difference from
\cite{Grundy2011} the diameter 
range for Borasisi (primary component) 
is 129--205 km. Our 
solution for the Borasisi-Pabu system
is $163_{-66}^{+32}$ km and the derived density
$2.1_{-1.2}^{+2.6}$ g cm$^{-3}$ (see Section~\ref{binaries}).
Our new estimate for the primary component is $126_{-51}^{+25}$ km (Table~\ref{bin}).

\begin{figure*}
   \includegraphics[width=17cm]{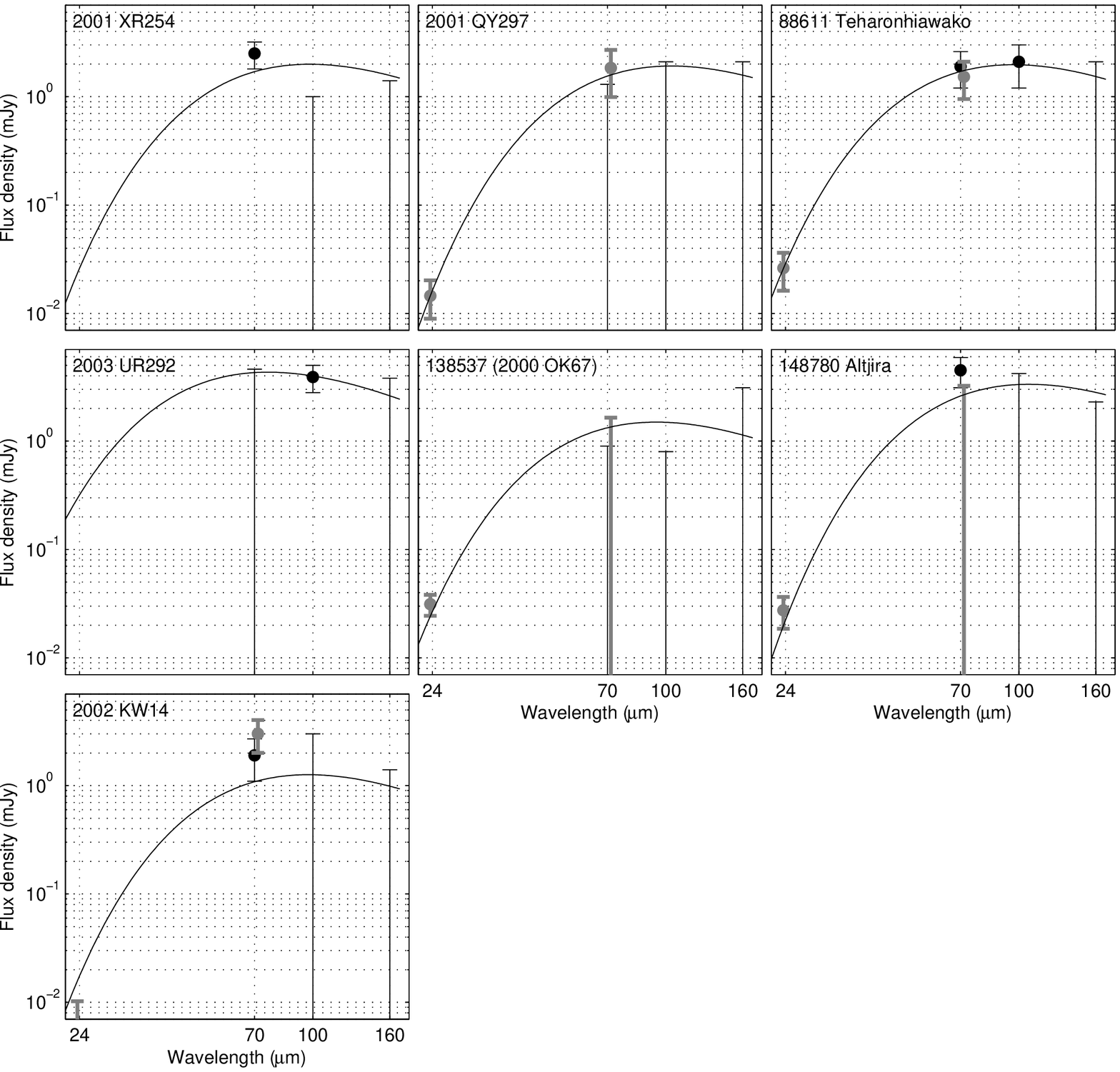} 
   \caption{SEDs calculated for remodeled targets from \cite{Vilenius2012} (2012). \emph{Spitzer} data
   (gray points at 24 and 71 $\mu m$) have been reduced to \emph{Herschel} observing geometry.}
   \label{fits_remodeled}
\end{figure*}

We have remodeled Teharonhiawako (from \cite[2012]{Vilenius2012}) with updated
\emph{Spitzer}/MIPS flux densities 
given in Table~\ref{table_Spitzerobs}.
The updated result gives a 24\% larger size and 34\% smaller albedo
(See Fig.~\ref{fits_remodeled} and Tables~\ref{collection_colds}--\ref{collection}
for all results). Previously, MIPS data reduction gave upper limits only for 2001 QY$_{297}$.
After updated data reduction from both instruments
the solution of 2001 QY$_{297}$ is now
based on a floating-$\eta$ fit instead of a fixed-$\eta$ as was the case previously
in \cite{Vilenius2012} (2012). The new albedo estimate is lower, and the new size 
estimate is 15\% larger. Altjira, which has updated PACS flux densities, is now
estimated to be 29\% larger than in \cite[2012]{Vilenius2012}.
The dynamically hot CKBOs 1996 TS$_{66}$ and 2002 GJ$_{32}$, which have only
\emph{Spitzer} observations (\cite[2009]{Brucker2009}), have been remodeled (see
Table~\ref{collection}) after significant changes in flux densities.
In our new estimates target 2002 GJ$_{32}$ has low albedo and large size,
whereas the result of \cite{Brucker2009} (2009) was a smaller target with
moderately high albedo. Contrary, 1996 TS$_{66}$'s new size estimate is
smaller than the previous one, with higher albedo.

Due to the different treatment of upper limits
(Section~\ref{upplims}) the 
size estimates of 
2000 OK$_{67}$, 
2001 XR$_{254}$, 
2002 KW$_{14}$, and 2003 UR$_{292}$ have changed while input values in the
modeling are the same as in \cite{Vilenius2012} (2012)
(see Tables~\ref{collection_colds}-\ref{collection} and Fig.~\ref{fits_remodeled}). 
The authors of 
that work had ignored
all three upper limits of 2002 KW$_{14}$ to obtain a floating-$\eta$ fit for this target
but with the new treatment of upper limits there is no need to ignore any data.
Instead of a 319 km target with geometric albedo 0.08 the new solution gives a high geometric
albedo of 0.31 and a diameter of 161 km.
The only case where we have ignored one upper limit is 2000 OK$_{67}$, which has four
upper limits and was not detected by PACS. The upper limit at 160 $\mu m$ is an outlier
compared to the others at 70-100 $\mu m$ and therefore we do not assume that band to be
close to the detection limit (see the adopted solution in Fig.~\ref{fits_remodeled}).

\section{Sample results and discussion}
\label{discussions}
In planning the \emph{Herschel} observations we used a default assumption for geometric albedo
of 0.08. As seen in Tables~\ref{collection_colds}-\ref{collection}, almost all 
dynamically cold CKBOs and more than half 
of hot CKBOs have higher
albedos implying lower flux densities at far-infrared wavelengths. This has lead to the
moderate SNRs and several upper limit flux densities in our sample.
The frequency of binaries among the cold CKBOs is high due to the selection process
of Herschel targets (see Section 2.1). We use this sample of cold
CKBOs, affected by the binarity bias, in the debiasing procedure of their size distribution
because of the very small number of non-binaries available.
In the analysis of 
sample properties of CKBOs we sometimes use 
a restricted sample, which we call ``regular'' CKBOs, where 
dwarf planets (Quaoar, Varuna, Makemake) and Haumea
family members (Haumea and 2002 TX$_{300}$) have been excluded.
All five targets mentioned are dynamically hot so that no cold CKBOs
are excluded when analysing the ``regular CKBOs`` sample.

\begin{table*}
\centering
\caption{Adopted physical properties of cold classical TNOs.}
\begin{tabular}{llcccllcl}
\hline
  Target                  &   & i ($\degr$) & a (AU) & D (km) & p$_V$             & $\eta$                 & No. of bands    & Reference \\
\hline
(2001 QS$_{322}$)                &   & 0.2 & 44.2 & $186_{-24}^{+99}$ & $0.095_{-0.060}^{+0.531}$ & (fixed) & 5 & This work \\
\noalign{\smallskip}
66652 Borasisi (1999 RZ$_{253}$) & B & 0.6 & 43.9 & $163_{-66}^{+32}$ & $0.236_{-0.077}^{+0.438}$ & $0.77_{-0.47}^{+0.19}$ & 5 & This work \\
\noalign{\smallskip}
(2003 GH$_{55}$)                 &   & 1.1 & 44.0 & $178_{-56}^{+21}$ & $0.150_{-0.031}^{+0.182}$ & (fixed)                & 3 & This work \\
\noalign{\smallskip}
(2001 XR$_{254}$)                & B & 1.2 & 43.0 & $221_{-71}^{+41}$ & $0.136_{-0.044}^{+0.168}$ & (fixed)                & 3 & (*) \cite{Vilenius2012} (2012) \\
\noalign{\smallskip}
275809  (2001 QY$_{297}$)        & B & 1.5 & 44.0 & $229_{-108}^{+22}$ & $0.152_{-0.035}^{+0.439}$ & $1.52_{-0.92}^{+0.22}$ & 5 & (*) \cite{Vilenius2012} (2012) \\
\noalign{\smallskip}
(2002 VT$_{130})$                & B & 1.2 & 42.7 & $324_{-68}^{+57}$  & $0.097_{-0.049}^{+0.098}$ & 1.20$\pm$0.35          & 2 & \cite{Mommert2013} \\
\noalign{\smallskip}
(2001 QB$_{298})$                &   & 1.8 & 42.6 & $196_{-53}^{+71}$  & $0.167_{-0.082}^{+0.162}$ & 1.20$\pm$0.35          & 2 & \cite{Mommert2013} \\
\noalign{\smallskip}
(2001 RZ$_{143}$)                & B & 2.1 & 44.4 & $140_{-33}^{+39}$ & $0.191_{-0.045}^{+0.066}$ & $0.75_{-0.19}^{+0.23}$ & 5 & \cite{Vilenius2012} (2012) \\
\noalign{\smallskip}
(2002 GV$_{31}$)                 &   & 2.2 & 43.9 & $<$180            & $>$0.19                   & (fixed)                & 3 & (*) \cite{Vilenius2012} (2012) \\
\noalign{\smallskip}
79360 Sila                       & B & 2.2 & 43.9 & $343 \pm 42$ & $0.090_{-0.017}^{+0.027}$ & $1.36_{-0.19}^{+0.21}$ & 5 & \cite{Vilenius2012} (2012) \\
\noalign{\smallskip}
(2003 QA$_{91}$)                 & B & 2.4 & 44.5 & $260_{-36}^{+30}$ & $0.130_{-0.075}^{+0.119}$ & $0.83_{-0.15}^{+0.10}$ & 5 & This work \\
\noalign{\smallskip}
88611 Teharonhiawako             & B & 2.6 & 44.2 & $220_{-44}^{+41}$ & $0.145_{-0.045}^{+0.086}$ & $1.08_{-0.28}^{+0.30}$ & 5 & (*) \cite{Vilenius2012} (2012) \\
\noalign{\smallskip}
(2005 EF$_{298}$)                & B & 2.9 & 43.9 & $174_{-32}^{+27}$ &  $0.16_{-0.07}^{+0.13}$    & (fixed)                & 3 & \cite{Vilenius2012} (2012) \\
\noalign{\smallskip}
(2003 QR$_{91}$)                 & B & 3.5 & 46.6 & $280_{-30}^{+27}$ & $0.054_{-0.028}^{+0.035}$  & $1.20_{-0.12}^{+0.10}$ & 2 & This work \\
\noalign{\smallskip}
(2003 WU$_{188}$)                & B & 3.8 & 44.3 & $<$220 &  $>$0.15    & (fixed)                & 3 & This work \\
\hline
\end{tabular}
\label{collection_colds}
\tablefoot{'B' indicates a known binary system (\cite{Noll2008}, \cite{Noll2009}, \cite{Benecchi2013})
and the diameter given is the area-equivalent system diameter.
(*) marks a target remodeled in this work using input data from the reference.}
\end{table*}

\begin{table*}
\centering
\caption{Adopted physical properties of hot classical TNOs.}
\begin{tabular}{llrcclccl}
\hline
  Target                  &   & i ($\degr$) & a (AU) & D (km)     & p$_V$                        & $\eta$      & No. of bands & Reference \\
\hline
\noalign{\smallskip}
2002 KX$_{14}$  &   &  0.4 & 38.9 & $455 \pm 27$        &  $0.097_{-0.013}^{+0.014}$    & $1.79_{-0.15}^{+0.16}$ & 5 & \cite{Vilenius2012} (2012) \\
\noalign{\smallskip}
2001 QT$_{322}$ &   &  1.8 & 37.2 & $159_{-47}^{+30}$   &  $0.085_{-0.052}^{+0.424}$    & (fixed)                & 5 & This work \\
\noalign{\smallskip}
2003 UR$_{292}$ &   &  2.7 & 32.6 & $136_{-26}^{+16}$   &  $0.105_{-0.033}^{+0.081}$    & (fixed)                & 3 & (*) \cite{Vilenius2012} (2012)     \\
\noalign{\smallskip}
1998 SN$_{165}$ &   &  4.6 & 38.1 & $393_{-38}^{+39}$        &  $0.060_{-0.013}^{+0.019}$    & (fixed) & 4 & This work \\
\noalign{\smallskip}
2000 OK$_{67}$  &   &  4.9 & 46.8 & $164_{-45}^{+33}$   &  $0.169_{-0.052}^{+0.159}$    & (fixed)                & 5 & (*) \cite{Vilenius2012} (2012) \\
\noalign{\smallskip}
2001 QD$_{298}$ &   &  5.0 & 42.7 & $233_{-63}^{+27}$   &   $0.067_{-0.014}^{+0.062}$       & (fixed) & 5 & This work \\
\noalign{\smallskip}
148780 Altjira  & B &  5.2 & 44.5 & $331_{-187}^{+51}$   &  $0.0430_{-0.0095}^{+0.1825}$    & $1.62_{-0.83}^{+0.24}$ & 5 & (*) \cite{Vilenius2012} (2012) \\
\noalign{\smallskip}
1996 TS$_{66}$  &   &  7.3 & 44.2 & $159_{-46}^{+44}$   & $0.179_{-0.070}^{+0.173}$     & $0.75_{-0.27}^{+0.21}$ & 2 & (*) \cite{Brucker2009} (2009) \\ 
\noalign{\smallskip}
50000 Quaoar    & B &  8.0 & 43.3 & $1074 \pm 38$       & $0.127_{-0.009}^{+0.010}$            & $1.73\pm 0.08$ & 8 & \cite{Fornasier2013} (2013) \\ 

\noalign{\smallskip}
2002 KW$_{14}$  &   &  9.8 & 46.5 & $161_{-40}^{+35}$   &   $0.31_{-0.094}^{+0.281}$       & (fixed)                & 5 & (*) \cite{Vilenius2012} (2012) \\
\noalign{\smallskip}
2002 GJ$_{32}$  &   & 11.6 & 44.1 & $416_{-78}^{+81}$   &  $0.035_{-0.011}^{+0.019}$     & $2.05_{-0.36}^{+0.38}$  & 2 & (*) \cite{Brucker2009} (2009) \\ 
\noalign{\smallskip}
2001 KA$_{77}$  &   & 11.9 & 47.3 & $310_{-60}^{+170}$  & $0.099_{-0.056}^{+0.052}$    & $2.52_{-0.83}^{+0.18}$ & 5 & \cite{Vilenius2012} (2012) \\
\noalign{\smallskip}
19521 Chaos     &   & 12.0 & 46.0 & $600_{-130}^{+140}$ &  $0.050_{-0.016}^{+0.030}$    & $2.2_{-1.1}^{+1.2}$    & 4 & \cite{Vilenius2012} (2012) \\
\noalign{\smallskip}
2002 XW$_{93}$  &   & 14.3 & 37.6 & $565_{-73}^{+71}$   &  $0.038_{-0.025}^{+0.043}$    & $0.79_{-0.24}^{+0.27}$ & 3 & \cite{Vilenius2012} (2012) \\
\noalign{\smallskip}
20000 Varuna    &   & 17.2 & 43.0 & $668_{-86}^{+154}$ & $0.127_{-0.042}^{+0.040}$    & $2.18_{-0.49}^{+1.04}$ & 3 & \cite{Lellouch2013} (2013) \\ 
\noalign{\smallskip}
2002 MS$_4$     &   & 17.7 & 41.7 & $934 \pm 47$        &  $0.051_{-0.022}^{+0.036}$    & $1.06 \pm 0.06$        & 5 & \cite{Vilenius2012} (2012) \\
\noalign{\smallskip}
2005 RN$_{43}$  &   & 19.2 & 41.8 & $679_{-73}^{+55}$   &   $0.107_{-0.018}^{+ 0.029}$  & (fixed)                & 3 & \cite{Vilenius2012} (2012)    \\
\noalign{\smallskip}
2002 UX$_{25}$  & B & 19.4 & 42.8 & $697 \pm 35$        & $0.107 \pm 0.010$            & $1.07_{-0.05}^{+0.08}$        & 8 & \cite{Fornasier2013} (2013) \\ 
\noalign{\smallskip}
% 2003 MW$_{12}$          
174567 Varda    & B & 21.5 & 45.6 & $792_{-84}^{+91}$ & $0.102_{-0.020}^{+0.024}$ & $0.84_{-0.22}^{+0.28}$ & 3 & This work \\
\noalign{\smallskip}
2004 GV$_9$     &   & 22.0 & 41.8 & $680 \pm 34$        & $0.0770_{-0.0077}^{+0.0084}$ & $1.93_{-0.07}^{+0.09}$ & 5 & \cite{Vilenius2012} (2012) \\
\noalign{\smallskip}
1999 RY$_{215}$ &   & 22.2 & 45.5 & $263_{-37}^{+29}$   & $0.0388_{-0.0065}^{+0.0122}$ & (fixed)                & 3 & This work \\
\noalign{\smallskip}
120347 Salacia  & B & 23.9 & 42.2 & $901 \pm 45$        &  $0.044_{-0.004}^{+0.004}$    & $1.16 \pm 0.03$        & 8 & \cite{Fornasier2013} (2013) \\
\noalign{\smallskip}
2002 AW$_{197}$ &   & 24.4 & 47.2 & $768_{-38}^{+39}$   &  $0.112_{-0.011}^{+0.012}$    & $1.29_{-0.10}^{+0.13}$ & 5 & This work \\
\noalign{\smallskip}
2005 UQ$_{513}$ &   & 25.7 & 43.5 & $498_{-75}^{+63}$ &  $0.202_{-0.049}^{+0.084}$    & (fixed) & 3 & This work \\
\noalign{\smallskip}
2002 TX$_{300}$ &   & 25.8 & 43.5 & $286 \pm 10$        &   $0.88_{-0.06}^{+0.15}$       & $1.15_{-0.74}^{+0.55}$     & occultation & \cite{Elliot2010} (2010),  \\
                &   &      &      &                     &                                &                            & +3          & \cite{Lellouch2013} (2013) \\
\noalign{\smallskip}
2004 PT$_{107}$ &   & 26.1 & 40.6 & $400_{-51}^{+45}$ &  $0.0325_{-0.0066}^{+0.0111}$    & (fixed)     & 3 & This work \\
\noalign{\smallskip}
2002 GH$_{32}$  &   & 26.7 & 41.9 & $<$\,180            & $>$\,0.13                    & (fixed)                & 3 & This work \\
\noalign{\smallskip}
136108 Haumea   & B & 28.2 & 43.1 & $1240_{-59}^{+69}$ & $0.804_{-0.095}^{+0.062}$  & $0.95_{-0.26}^{+0.33}$     & 3             & \cite[2013]{Fornasier2013} \\
\noalign{\smallskip}
136472 Makemake &   & 29.0 & 45.5 & $1430 \pm 9$       & 0.77$\pm$0.03                & $2.29_{-0.40}^{+0.46}$     & occultation & \cite{Ortiz2012} (2012),  \\
                &   &      &      &                    &                              &                            & +3          & \cite{Lellouch2013} (2013) \\
\noalign{\smallskip}
2001 QC$_{298}$  & B  & 30.6 & 46.3 & $303_{-30}^{+27}$ & $0.061_{-0.017}^{+0.027}$ & $0.985_{-0.095}^{+0.084}$ & 2 & This work \\
\noalign{\smallskip}
2004 NT$_{33}$ &   & 31.2  & 43.5  & $423_{-80}^{+87}$ & $0.125_{-0.039}^{+0.069}$    & $0.69_{-0.32}^{+0.46}$ & 3 & This work \\
\noalign{\smallskip}
2004 XA$_{192}$ &   & 38.1 & 47.4  & $339_{-95}^{+120}$ & $0.26_{-0.15}^{+0.34}$      & $0.62_{-0.49}^{+0.79}$ & 3 & This work \\
\noalign{\smallskip}
\hline
\end{tabular}
\label{collection}
\tablefoot{'B' indicates a known binary system and the diameter given is the area-equivalent system diameter.
(*) marks a target remodeled in this work.}
\end{table*}

\subsection{Measured sizes}
The diameter estimates in the "regular CKBO'' sample are ranging from 136 km of 
2003 UR$_{292}$ up to 934 km of 2002 MS$_4$.
The 
not detected targets (2002 GV$_{31}$, 2003 WU$_{188}$ and 2002 GH$_{32}$)
may be smaller than 
2003 UR$_{292}$. 
Dynamically cold targets in our measured sample
are limited to diameters of 100-400 km whereas hot CKBOs have a much wider size distribution
up to sizes of $\sim$900 km in our measured "regular CKBO'' sample and up to 1430 km
when dwarf planets are included.

\begin{figure}
   \resizebox{\hsize}{!}{\includegraphics{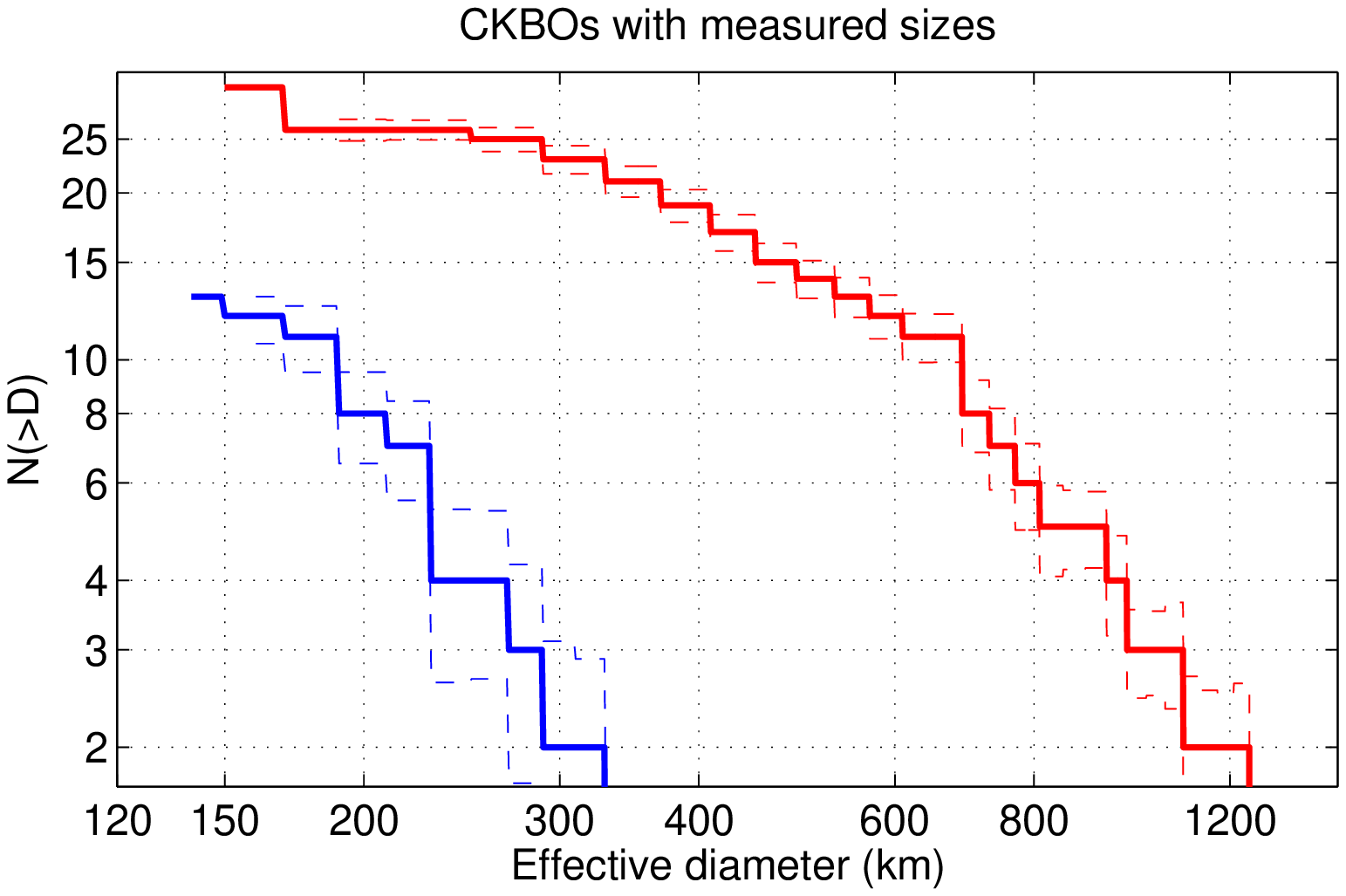}}
   \caption{Cumulative size distribution of all measured dynamically hot (red, upper)
    classicals from Table~\ref{collection} and dynamically cold (blue, lower) classicals from
    Table~\ref{collection_colds}. 
    The dotted lines are uncertainties obtained by a Monte Carlo technique
    where the sizes of targets are varied using their measured error bars, and a set of
    size distributions are created. The standard deviation of these size distributions
    is calculated at each size bin.
    }
   \label{diameter_cumulative}
\end{figure}

We show the cumulative size distribution $N(>$$D)\propto D^{1-q}$
of 
hot classicals (from Table~\ref{collection}) and 
cold classicals (from Table~\ref{collection_colds})
in Fig.~\ref{diameter_cumulative}
\footnote{Note that in \cite{Vilenius2012} (2012) the authors used a different definition: $N(>D) \propto D^{-q}$,
but that notation differs from most of the literature.}.
In the measured, biased, size distribution of hot CKBOs we can distinguish
three regimes for the power law slope:
$100$$<$$D$$<$$300\ \mathrm{km}$, 
$400$$<$$D$$<$$600\ \mathrm{km}$ and $700$$<$$D$$<$$1300\ \mathrm{km}$.
The slope parameters
for 
the latter two regimes are
$q$$\approx$2.0 and $q$$\approx$4.0. In the small-size regime there are not enough
targets in different size bins to derive a reliable slope. The measured, biased, cold CKBO sample 
gives a slope of $q$$\approx$4.3 in the size range 200$<$$D$$<$300 km. The debiased
size distribution slopes are given in Section~\ref{CFEPSsizes}.

\subsection{Measured geometric albedos}
\label{albedodiscussion}
Haumea family members and many dwarf planets have very high geometric albedos.
The highest-albedo regular CKBO is 2002 KW$_{14}$ with
$p_\mathrm{V}$$=$0.31 and the darkest object is 
2004 PT$_{107}$ with $p_\mathrm{V}$$=$0.0325,
both dynamically hot.
Among dynamically
cold objects geometric albedo is between Sila's $p_\mathrm{V}$$=$0.090
and Borasisi's $p_\mathrm{V}$$=$0.236.

\begin{figure}
   \resizebox{\hsize}{!}{\includegraphics{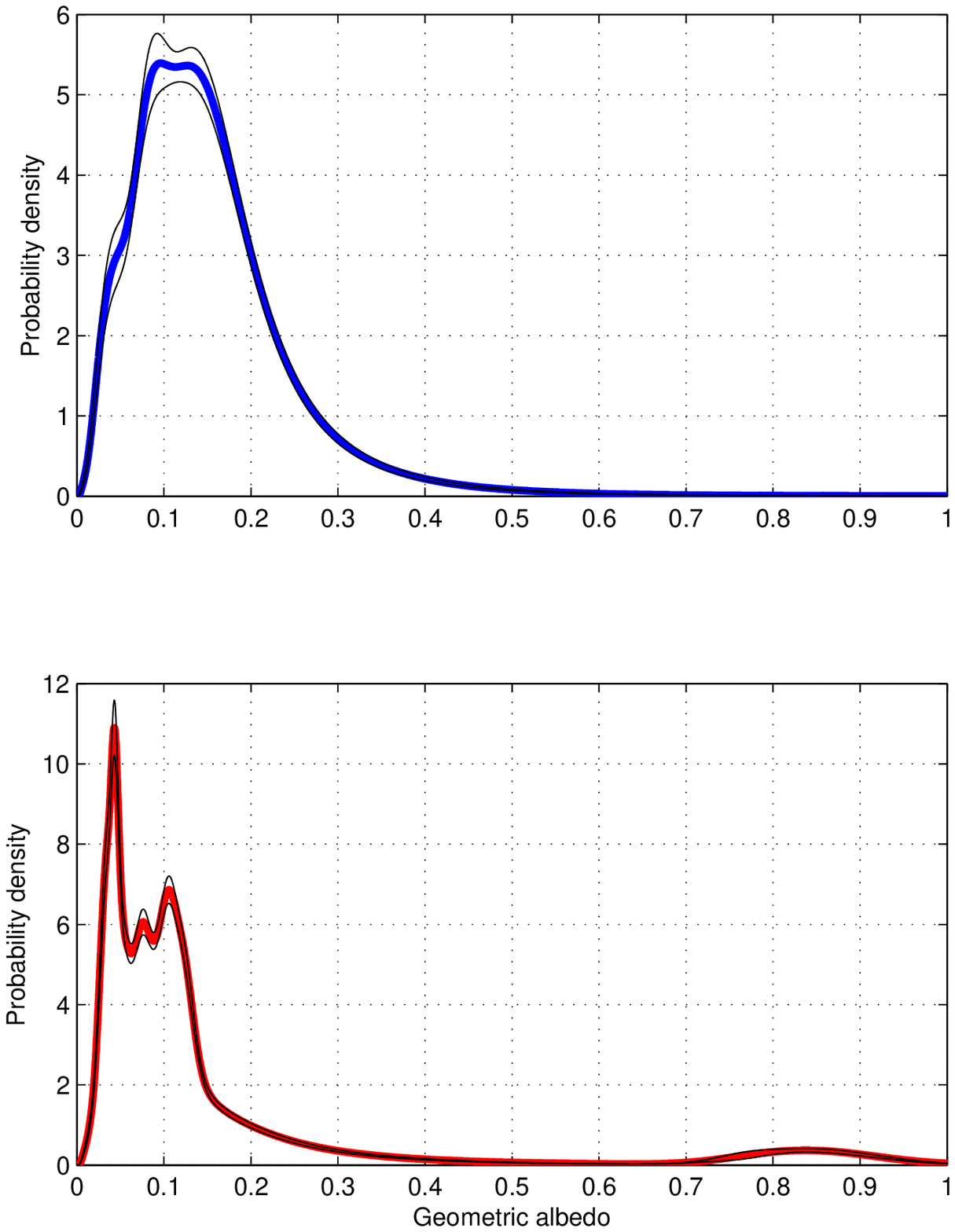}}
   \caption{Distribution of measured geometric albedos: upper panel for cold CKBOs
   and lower panel for hot CKBOs. The thin lines show the standard deviation of all
   probability density distributions, where each of the distributions has been determined
   with one target excluded, each target having been excluded once.}
   \label{albedo_histo}
\end{figure}

The sub-sample of cold CKBOs
are lacking low-albedo objects compared to the hot sub-sample.
Fig.~\ref{albedo_histo} shows probability density functions
constructed from the measured geometric albedos and their asymmetric error bars
using the technique described in detail in \cite{Mommert2013}.
The probability density for each individual target is assumed to follow a lognormal distribution,
whose scale parameter is calculated using the upper and lower uncertainties given
for the measured geometric albedo.
The 
median geometric albedo of the combined probability density (Fig.~\ref{albedo_histo}) of
cold classicals 
is $0.14_{-0.07}^{+0.09}$,  
of
regular hot CKBOs 
$p_\mathrm{V}$$=$$0.085_{-0.045}^{+0.084}$, and of
all hot CKBOs including
dwarf planets and Haumea family the median is $p_\mathrm{V}$$=$$0.10^{+0.16}_{-0.06}$. 
These medians
are compatible with 
averages
obtained from smaller sample sizes:
$0.17$$\pm$0.04 for cold CKBOs and $0.11$$\pm$0.04 for hot CKBOs in \cite{Vilenius2012} (2012)
but the difference between the 
dynamically cold and hot
sub-samples 
is smaller than previously reported.

Of the other dynamical classes, the Plutinos have an average albedo of 0.08$\pm$0.03
(\cite[2012]{Mommert2012}), scattered disk objects have 0.112 (\cite[2012]{SantosSanz2012}), detached objects have 0.17
(\cite[2012]{SantosSanz2012}), gray Centaurs have 0.056 (\cite{Duffard2013}) and red Centaurs 0.085
(\cite{Duffard2013}). Dynamically hot classicals have a similar average albedo as Plutinos and red Centaurs
whereas the average albedo of cold CKBOs is closer to the detached objects.

\subsection{Debiased size distributions}
\label{CFEPSsizes}
The measured size distributions are affected by biases: the radiometric
method has a detection limit, and the measured sample is not representative of
all those targets which could have been detectable in principle. For the
debiasing we use a synthetic model of outer Solar System objects by the
Canada-French Ecliptic Plane Survey (CFEPS, \cite{Petit2011}), which is based
on well-calibrated optical surveys. CFEPS provides $H_g$ magnitudes and orbital 
parameters of more than 15000 cold
CKBOs and 35000 hot synthetic CKBOs.
We perform a two-stage debiasing of the measured size distribution (see Appendix A for details)
and derive slope parameters. We have constructed a model of the detection
limit of \emph{Herschel} observations, which depends on objects' sizes, albedos and distances.
This model is used in the first stage of debiasing. In the second stage we debias
the size distribution in terms of how the distribution of $H_g$s of the measured
targets are related to the $H_g$ distribution of the synthetic sample of those objects,
which would have been detectable.

CFEPS has synthetic objects to the limit of $H_g$=8.5. All cold CKBOs
in our measured sample have $H_g<$7.5 and all hot CKBOs have $H_g<$8.0.
Therefore, these limits are first applied to the CFEPS sample before
debiasing the size distributions. Since all of the measured hot CKBOs are in the
inner or main classical belts, we exclude the outer CKBOs of CFEPS in the debiasing.
Furthermore, we have excluded a few
measured targets which are outside the orbital elements space of CFEPS
objects, or which are close to the limit of dynamically cold/hot CKBOs,
to avoid contamination from one sub-population to the other.

In translating the optical absolute magnitude of simulated CFEPS objects into sizes,
a step needed in the debiasing (Appendix~\ref{debiasing}), we use the measured
albedo probability densities (Fig. ~\ref{albedo_histo}) in a statistical way.
Our measured dynamically hot CKBOs cover the relevant heliocentric distance range
of inner and main classical belt CFEPS objects. While our measured sample of cold
CKBOs is limited to 38$<$$r$$<$45 AU we assume
that the shape of the albedo distribution applies also to more distant cold CKBOs.
Although there is an optical
discovery bias prefering high-$p_V$ objects
at large distances, 
the radiometric method has an opposite bias:
low-$p_V$ objects are easier to detect at thermal wavelengths than high-$p_V$ objects.
Among the radiometrically measured targets we do not find evidence of any
significant correlations (see Section~\ref{correlations})
between geometric albedo and orbital elements,
heliocentric distance at discovery time nor ecliptic latitude at discovery time.

\begin{figure}
   \resizebox{\hsize}{!}{\includegraphics{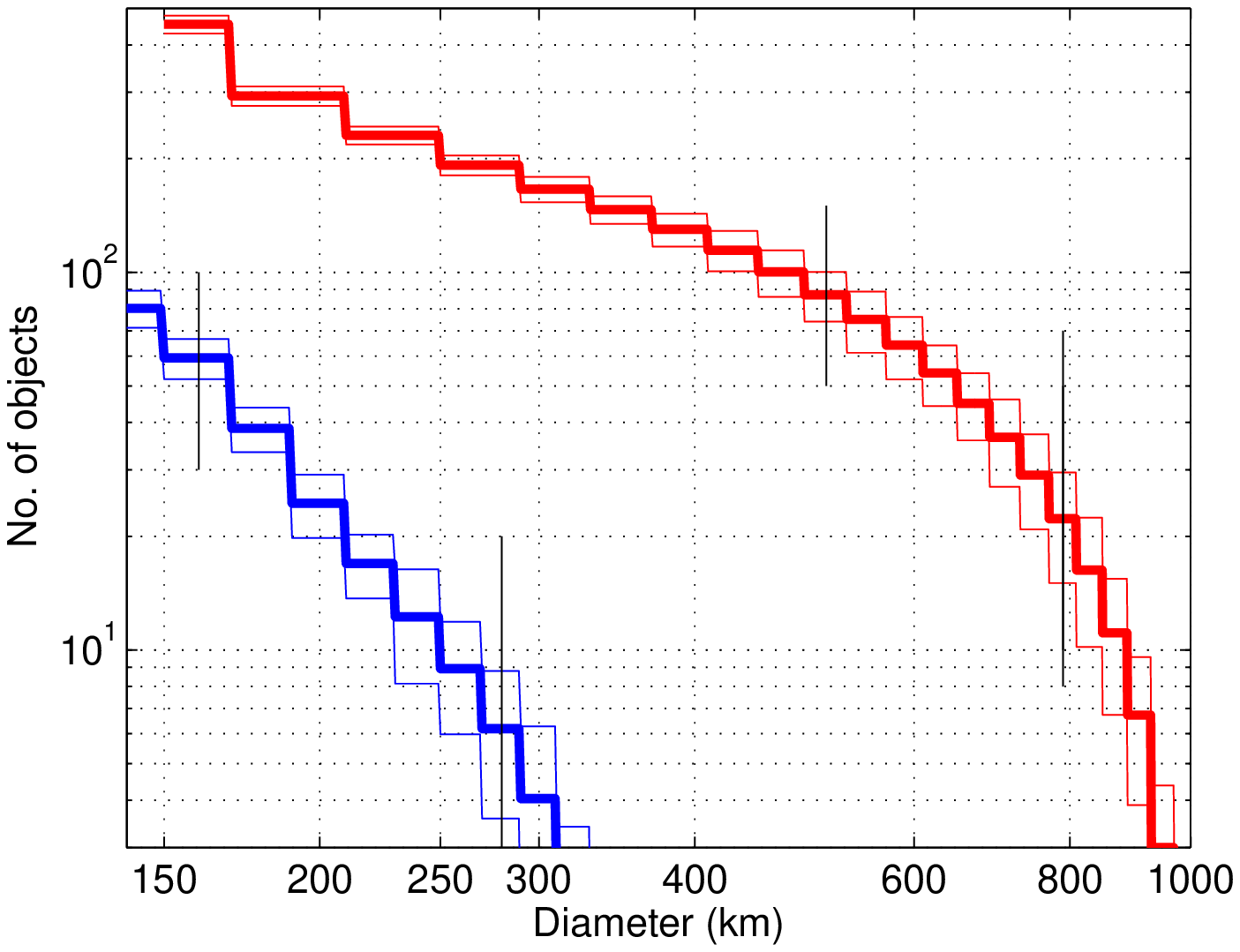}}
   \caption{Debiased size distributions (see text): hot CKBOs (red, upper) and cold CKBOs (blue, lower).
   The thin red and blue lines are the error bars of each size bin (bin size 20 km for
   dynamically cold and 40 km for dynamically hot CKBOs).
   The vertical lines mark the ranges for which slope parameters have been determined.}
   \label{sizes_debiased}
\end{figure}

The debiased size distributions are shown in Fig.~\ref{sizes_debiased}.
Our analysis of cold CKBOs gives a debiased slope of $q$=5.1$\pm$1.1 in the
range of effective diameters of 160$<$$D$$<$280 km. 
In the measured sample there are seven binaries and three non-binaries in
this size range.
For dynamically hot
CKBOs the slope is 
$q$=2.3$\pm$0.1 in the size range 100$<$$D$$<$500 km. The slope is
steepening towards the end tail of the size distribution and in the
size range 500$<$$D$$<$800 km we obtain a slope parameter of $q$=4.3$\pm$0.9.
When comparing the slopes of the cold and hot sub-populations it should be
noted that for the cold subsample we are limited to the largest objects
and the maximum size of cold CKBOs is smaller than that of hot CKBOs.

Size distribution is often derived from the LF using simplifying assumptions about
common albedo and distance. 
\cite{Fraser2010} (2010) have derived a LF based slope for dynamically cold
objects ($i$$<$5 deg, 38$<$$r$$<$55 AU): $q$=5.1$\pm$1.1, which is 
well compatible with
our value from a debiased
measured size distribution. For dynamically hot CKBOs \cite{Fraser2010} (2010)
derived 
two slopes depending on the distance of objects. For dynamically hot objects
with 38$<$$r$$<$55 AU and $i$$>$5 deg: $q$$=$2.8$\pm 1.0$
and for a combined sample of these hot and ``close'' objects (30$<$$r$$<$38 AU):
$q$$=$3.0$\pm 0.6$. Both of the LF based results are
compatible within the given uncertainties with our estimate of $q$=2.3$\pm$0.1.

\subsection{Beaming factors}
The temperature distribution over an airless object affects 
the observed SED shape. 
In the NEATM model temperature is adjusted by the beaming factor $\eta$ as explained in
Section~\ref{model}. For CKBOs, the PACS bands are close to the thermal peak of the SED whereas MIPS
provides also data from the short-wavelength part of the SED.
Therefore,
in order to determine a reliable estimate for the average beaming
factor of classical TNOs we select those solutions which are based on
detections with both PACS and MIPS and detected in at least three bands.
Furthermore, we require that the MIPS 24 $\mu m$ band has been detected because it constrains
the overall shape of the SED making inferences based on those results more reliable. 
There is a large scatter of beaming factors among CKBOs spanning the full
range of $0.6$$<$$\eta$$<$$2.6$.
There are
five cold CKBOs and eight hot CKBOs with floating-$\eta$ solutions fulfilling 
the above mentioned criteria.
The averages of the two subpopulations do not differ much compared to the standard deviations.
The average beaming factor of 13 cold and hot CKBOs is $\eta$$=$1.45$\pm$0.46 and the median is 1.29.
This average is very close to 
the previous average based on eight targets: $\eta$$=$1.47$\pm$0.43 (\cite[2012]{Vilenius2012}).
The new average $\eta$ is compatible with the default value of
1.20$\pm$0.35 for fixed $\eta$ fits as well as with averages of other dynamical classes:
seven Plutinos have the average $\eta$$=$$1.11_{-0.19}^{+0.18}$ (\cite[2012]{Mommert2012}) and
seven scattered and detached objects have $\eta$$=$1.14$\pm$0.15
(\cite[2012]{SantosSanz2012}).
Statistically, beaming factors of a large sample of TNOs from all dynamical classes
are dependent on heliocentric distance (\cite[2013]{Lellouch2013}).
Therefore, $\eta$ values are likely to differ due to different distances of the
populations in different dynamical classes.

\subsection{Correlations}
\label{correlations}
In the sample of measured objects we have checked possible 
correlations between geometric albedo $p_{\mathrm{V}}$, diameter $D$, orbital elements 
(inclination $i$, eccentricity $e$, semimajor axis $a$, perihelion distance 
$q$), beaming 
factor $\eta$, heliocentric distance at discovery time,
ecliptic latitude at discovery time,
visible spectral slope, as well as B-V, V-R and V-I colors. 
We use a modified form of the Spearman correlation test (\cite{Spearman1904}) taking into account
asymmetric error bars and small numbers statistics. 
The details of this method are
described in \cite{Peixinho2004} and \cite{SantosSanz2012} (2012, Appendix B.2.).
We consider correlation coefficient $\rho$ to show a 'strong correlation'
when $|\rho| \geq 0.6 $ and 'moderate correlation' when $0.3 \leq |\rho|<0.6$.
Our correlation method does not show any significant (confidence
on the presence of a correlation $>$3$\sigma$) correlations between any parameters
within the dynamically cold subpopulation with $N$=13 targets. Similarly,
when making the correlation analysis on the CKBOs according to the DES classification
(N=23) we do not find any significant correlations.

\subsubsection{Diameter and geometric albedo}
\label{corr_D}
There is a lack of large objects at small inclinations and of small objects at
high inclinations in our measured sample.
The latter are subject to discovery biases since many of the
surveys have been limited close to the ecliptic plane
($H_V$ and ecliptic latitude at discovery time show a moderate
anti-correlation in the sample of all radiometrically measured targets).
There is a strong size-inclination correlation when all targets are included 
(4.4$\sigma$), 
and 
a moderate correlation
if dwarf planets and Haumea family are excluded 
(3.9$\sigma$).
The strong correlation
within the sub-sample of hot CKBOs
reported by \cite{Vilenius2012} (2012) 
is only moderate with our larger number of targets, and it is no longer
significant (2.3-2.5$\sigma$).
\cite{Levison2001} (2001) found 
the
presumable size-inclination trend from the correlation between intrinsic brightness 
and inclination and showed that their result is unlikely to be caused by biases.
When observing with the radiometric techniques, there is a selection bias of
targets, which we estimate to have high enough flux density to be detectable. According to
Equation~(\ref{model_emission}) the observed
flux is approximately proportional to the projected area
and inversely proportional to the square of distance. A statistical study of 85 TNOs and Centaurs,
with partially overlapping samples with this work,
showed a strong ($\rho$$=$0.78, significance $>$8$\sigma$) correlation between diameter
and instantaneous heliocentric
distance (\cite[2013]{Lellouch2013}). 
Dynamically hot
CKBOs show a moderate correlation
between effective diameter and heliocentric distance at discovery time (3.2$\sigma$).
However, it is not significant
when analyzed without dwarf planets and Haumea family (the ''regular`` sub-sample).
A diameter/inclination correlation could appear if there is a correlation between diameter and distance
as well as between distance and inclination. Our analysis finds no evidence of a correlation
between inclination and heliocentric distance for the whole measured sample ($\rho$$=$0.20, significance 1.2$\sigma$)
or any of the sub-samples.
Therefore, we consider the correlation between diameter and inclination reported here not to be
caused by a selection bias.

There is 
a moderate ($\rho$$\approx$-0.5) anti-correlation between diameter and geometric albedo
among the ``regular'' CKBOs 
(3.4$\sigma$).
This correlation disappears
when the dwarf planets and Haumea family members are added, or when the ``regular'' CKBOs is divided into
its cold or hot components. Inclination may be a common variable,
which correlates both with diameter 
and tentatively with
albedo as explained in the following. There is a moderate anticorrelation
between inclination and albedo among the ``regular'' CKBOs, although it is not
considered significant 
(2.5$\sigma$).
This is probably caused
mainly by the cold CKBOs 
($\rho$=-0.51, 1.8$\sigma$, N=13)
and less by the ``regular'' hot CKBOs
($\rho$=-0.17, 0.8$\sigma$, N=26).
When combining the
significant diameter/inclination correlation (3.9$\sigma$) with a tentative albedo/inclination correlation
this combination may explain
the moderate diameter/albedo anti-correlation we observe in our ``regular'' CKBOs sample. Therefore,
we do not confirm the finding of \cite{Vilenius2012} (2012) about an anti-correlation between size and albedo
within the classical TNOs as it is probably due to a bias.
The anti-correlation between size and albedo was not observed in Plutinos
(\cite[2012]{Mommert2012}), which do not show any correlation between size and albedo, nor with 
scattered/detached-disc objects which show a positive correlation
instead (\cite[2012]{SantosSanz2012}).

We find no evidence of other correlations with orbital elements or colors involving size or geometric albedo.

\subsubsection{Other correlations}
\label{corr_other}
CKBOs are
known 
to have an anti-correlation between surface color/spectral slope and orbital inclination
(\cite{Trujillo2002}, \cite[2002]{Hainaut2002}).
In our measured sample 
a moderate
correlation exists for the whole sample (3.2$\sigma$) but
is not significant for the ``regular'' CKBOs (2.0$\sigma$), which do not
include dwarf planets and Haumea family members. 
We do not find any correlations of
the B-V, V-R and V-I colors with 
parameters other than spectral slope.

The apparent $H_{\mathrm{V}}$ vs $i$ anti-correlation in our target sample mentioned in Section~\ref{Targetsample}
is moderate and  
significant 
for the whole sample 
(3.9$\sigma$)
as well as for the
hot sub-population (3.1$\sigma$), 
but less significant on the ``regular'' hot CKBOs sub-sample (2.5$\sigma$).

\subsection{Binaries}
\label{binaries}
\begin{figure}
   \resizebox{\hsize}{!}{\includegraphics{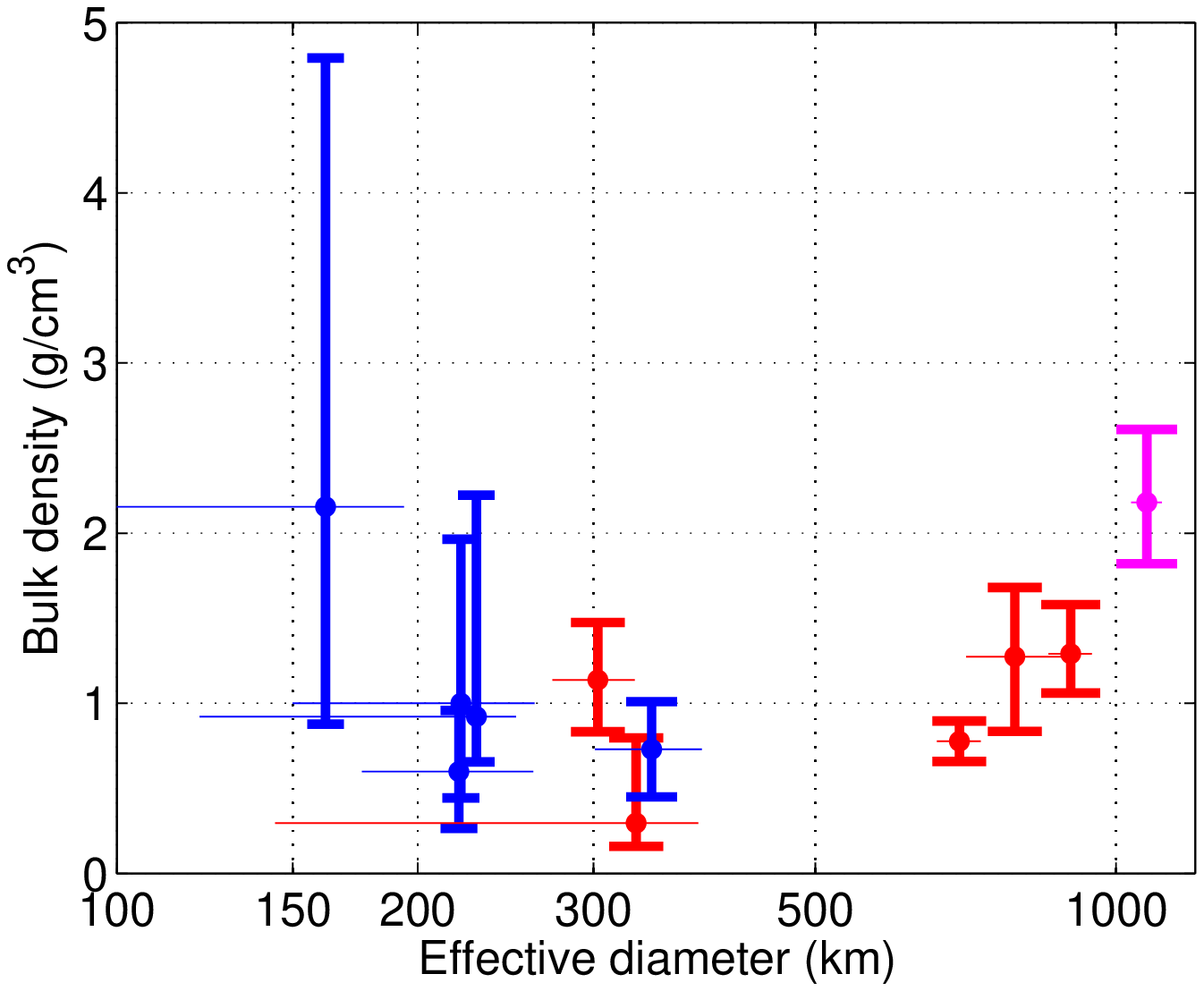}}
   \caption{Bulk densities of classical TNOs. Blue marks cold CKBOs,
   red hot CKBOs, and magenta dwarf planet Quaoar.}
   \label{bulkdensity_fig}
\end{figure}

In deriving bulk densities of binary systems, whose effective diameter $D$ has been
determined by the radiometric method, we assume that the primary and secondary
components i) are spherical, and ii) have equal albedos. 
A known brightness difference between the two components $\Delta V$ 
can be written as 
$k=10^{-0.2\Delta\mathrm{V}}=\frac{D_2}{D_1} / l$, where
$D_1$ and $D_2$ are the diameters of the primary and the secondary component and
$l=\sqrt{p_\mathrm{V1} / p_\mathrm{V2}}$ (components' geometric albedos $p_\mathrm{V1}$ and
$p_\mathrm{V2}$). 
The radiometric (area-equivalent) effective diameter of the system is $D^2 = D_1^2 + D_2^2$ and
the ``volumetric diameter'' is
$D_\mathrm{Vol} = \frac{\left( 1+\left( kl \right) ^3 \right)^{1/3}}{\sqrt{1+\left( kl \right)^2}} D$, which is then used
in calculating mass densities: $\frac{6 m}{\pi D_\mathrm{Vol}^3}$ with the usual assumption that l equals unity.
The new radiometric mass density estimates of Borasisi, Varda % 2003 MW$_{12}$ 
and 2001 QC$_{298}$, and
updated (see Section~\ref{resultscomp})
densities of Teharonhiawako,
Altjira, 2001 XR$_{254}$, and 2001 QY$_{297}$ are given in Table~\ref{bin_dens} and shown in Fig.~\ref{bulkdensity_fig}.
When $\Delta V$ is small, 
the density estimate 
does not change to significantly higher densities by changing
the assumed ratio of geometric albedos
unless the albedo contrast between the
primary and the secondary was extreme. The sizes of the binary components (Table~\ref{bin}) for
$<$400 km objects are not significantly different from each other. If we make the assumption that $D_1=D_2$
and determine densities and relative albedos we get densities close to those in Table~\ref{bin_dens}
for the $<$400 km objects and albedo ratios of 1.1-1.9.

The new 
density estimates
of four targets are lower than those determined by \cite{Vilenius2012} (2012). The reason
for the large change in density estimates is the \emph{Spitzer} flux update of three of the targets
and a different technique of treating upper limits in the cases of Teharonhiawako, Altjira and 2001 XR$_{254}$.
Our assumption that the objects are spherical may give too low density estimates for elongated objects.
The relatively large light curve variation of 2001 QY$_{297}$ of $\sim$0.5 mag (\cite{Thirouin2012} 2012)
suggests a shape effect whereas the light curve amplitude of Altjira is not well known and is
probably $<$0.3 mag (\cite{Sheppard2007}). Lower density limits can be derived based on rotational
properties but the period is not known for these two targets. Densities lower than that of water ice
have been reported for TNOs in the literature (e.g. \cite{Stansberry2006}).
The density of a sphere of pure water ice under self-compression is slightly less than 1 g cm$^{-3}$
and 
porosity at micro and macro scales reduces the bulk density.
Another common low-density ice is methane with a density of $\sim$0.5 g cm$^{-3}$. A statistical
study of TNOs from all dynamical classes shows that their surfaces are very porous
(\cite[2013]{Lellouch2013}) indicating that the material on the surface has a low density.
However, the low bulk densities
of Altjira and 
Teharonhiawako
reported here require significant 
porosities of 40-70\% for material densities of 0.5-1.0 g cm$^{-3}$. This would
indicate the presence of macroporosity, i.e. that the objects are rubble piles of 
icy pieces.

\begin{table*}
\centering
\caption{Density estimates of classical TNO binaries with a known mass. The primary and secondary
are assumed to have equal albedos and equal densities.}
\begin{tabular}{llcclc}
\hline
Target &  Adopted $\Delta$V\tablefootmark{a}  & Mass\tablefootmark{a}   & Bulk density / literature  & Reference & Bulk density / this work \\
       &  (mag)                               & ($\times 10^{18}$ kg)   & (g cm$^{-3}$)              &           & (g cm$^{-3}$) \\
\hline
\noalign{\smallskip}
Borasisi          & 0.45                     & $3.433 \pm 0.027$ & \ldots                            & \ldots & $2.1_{-1.2}^{+2.6}$ \\
\noalign{\smallskip}
2001 XR$_{254}$   & 0.43                     & $4.055 \pm 0.065$ & $1.4_{-1.0}^{+1.3}$               & \cite{Vilenius2012} (2012) & $1.00_{-0.56}^{+0.96}$ \\
\noalign{\smallskip}
2001 QY$_{297}$   & 0.20                     & $4.105 \pm 0.038$ & $1.4_{-1.3}^{+1.2}$               & \cite{Vilenius2012} (2012) & $0.92_{-0.27}^{+1.30}$ \\
\noalign{\smallskip}
Sila              & 0.12\tablefootmark{b}    & $10.84 \pm 0.22$\tablefootmark{b}  & 0.73$\pm$0.28    & \cite{Vilenius2012} (2012), (b) & \ldots \\
\noalign{\smallskip}
Teharonhiawako    & 0.70                     & $2.445 \pm 0.032$ & $1.14_{-0.91}^{+0.87}$            & \cite{Vilenius2012} (2012) & $0.60_{-0.33}^{+0.36}$  \\
\noalign{\smallskip}
Altjira           & 0.23                     & $3.986 \pm 0.067$ & $0.63_{-0.63}^{+0.68}$            & \cite{Vilenius2012} (2012) & $0.30_{-0.14}^{+0.50}$ \\
\noalign{\smallskip}
2002 UX$_{25}$    & $\sim$2.7\tablefootmark{g} & $\sim$125$\pm$3\tablefootmark{g}                    & 0.82$\pm$0.11 & \cite{Brown2013} & \\ 
\noalign{\smallskip}
% 2003 MW$_{12}$
Varda             & 1.45\tablefootmark{f}    & 265.1$\pm$3.9\tablefootmark{f}                        & \ldots             & \ldots & $1.27_{-0.44}^{+0.41}$ \\
\noalign{\smallskip}
2001 QC298        & 0.44\tablefootmark{f}    & 11.88$\pm$0.14\tablefootmark{f}  & \ldots             & \ldots & $1.14_{-0.30}^{+0.34}$ \\
\noalign{\smallskip}
Quaoar            & $5.6 \pm 0.2$\tablefootmark{e} & $1300-1400$\tablefootmark{c}  & $2.18_{-0.36}^{+0.43}$    & \cite{Fornasier2013} (2013) & \ldots \\
\noalign{\smallskip}
Salacia           & $2.372 \pm 0.060$\tablefootmark{d} & $436 \pm 11$\tablefootmark{d} & $1.29_{-0.23}^{+0.29}$ & \cite{Fornasier2013} (2013)  & \ldots \\
\noalign{\smallskip}
\hline
\end{tabular}
\label{bin_dens}
\tablefoot{{\bf References.}
\tablefoottext{a}{\cite{Grundy2011} unless otherwise indicated.} 
\tablefoottext{b}{\cite{Grundy2012}.}
\tablefoottext{c}{\cite{Fraser2012}.}
\tablefoottext{d}{\cite{Stansberry2012}.}
\tablefoottext{e}{\cite{Brown2007}.}
\tablefoottext{f}{Grundy et al. \emph{(in prep.)}}
\tablefoottext{g}{\cite{Brown2013}.}
}
\end{table*}

\begin{table}
\centering
\caption{Sizes of primary and secondary components assuming equal albedos and
spherical shapes of both components.}
\begin{tabular}{lcc}
\hline
Target &    Primary's size        & Secondary's size  \\
       &  $D_1$   (km)            & $D_2$  (km)         \\
\hline\noalign{\smallskip}
Borasisi           & $126_{-51}^{+25}$  & $103_{-42}^{+20}$ \\
\noalign{\smallskip}
2001 XR$_{254}$    & $171_{-55}^{+32}$  & $140_{-45}^{+26}$ \\
\noalign{\smallskip}
2001 QY$_{297}$    & $169_{-80}^{+16}$  & $154_{-73}^{+15}$  \\  
\noalign{\smallskip}
Sila               & $249_{-31}^{+30}$  & $236_{-29}^{+28}$  \\   
\noalign{\smallskip}
Teharonhiawako     & $178_{-36}^{+33}$  & $129_{-26}^{+24}$  \\   
\noalign{\smallskip}
Altjira            & $246_{-139}^{+38}$ & $221_{-125}^{+34}$  \\  
\noalign{\smallskip}
2002 UX$_{25}$     & 670$\pm$34         & 193$\pm$10 \\
\noalign{\smallskip}
% 2003 MW$_{12}$
Varda              & $705_{-75}^{+81}$  & $361_{-38}^{+42}$ \\
\noalign{\smallskip}
2001 QC298         & $235_{-23}^{+21}$  & $192_{-19}^{+17}$ \\
\noalign{\smallskip}
Quaoar \tablefootmark{a} & $1070\pm 38$      & $81\pm 11$      \\
\noalign{\smallskip}
Salacia            & $829\pm 30$ & $278\pm 10$   \\
\hline
\end{tabular}
\label{bin}
\tablefoot{\tablefoottext{a}{Quaoar's $D_1$ and $D_2$ from \cite{Fornasier2013} (2013).}}
\end{table}

\section{Conclusions}
\label{conclude}
The \emph{Herschel} mission and the cold 
phase of \emph{Spitzer} 
have ended. 
The next space mission capable of far-infrared observations of CKBOs will be in the next decade.
Occultations can provide very few new size estimates annually, and the capabilities of 
the \emph{Atacama Large Millimeter Array} (ALMA)
to significantly extend the sample of measured sizes of TNOs already presented
may be limited by its sensitivity\footnote{\cite{Moullet2011} estimated 500
TNOs to be detectable by ALMA, based on assumed albedos commonly used at that time.}.

In this work we have analysed 
18 classical TNOs to determine their sizes and albedos using the
radiometric technique and data from \emph{Herschel} and/or \emph{Spitzer}.
We have also re-analysed previously published 
targets, part of them with updated
flux densities. The number of CKBOs with size/albedo solutions in literature and this work is increased
to 44 
targets and additionally 
three targets have a diameter upper limit and albedo lower limit.
We have determined the mass density of three CKBOs and updated 
four
previous density estimates.
Our main conclusions are:
   \begin{enumerate}
    \item The dynamically cold CKBOs have higher geometric albedo ($0.14$)
          than the dynamically hot CKBOs ($0.085$ without dwarf planets and Haumea family, 0.10 including them),
          although the difference is not as great
          as reported by \cite{Vilenius2012} (2012). 
\item We do not confirm the general finding of \cite{Vilenius2012} (2012) that there is an anti-correlation between diameter
      and albedo among all measured CKBOs 
      as that analysis was based on a smaller number of targets.
    \item The cumulative size distributions of cold and hot CKBOs
          have been infered using a two-stage debiasing procedure.
          The characteristic size of cold CKBOs is smaller,
          which is compatible with the hypothesis that the cold sub-population may have formed
          at a larger heliocentric distance than the hot sub-population. The cumulative size distribution's slope parameters of hot CKBOs in the
          diameter range 100$<$$D$$<$500 km
          is $q$=2.3$\pm$0.1.
          Dynamically cold CKBOs have an infered slope of $q$$=$5.1$\pm$1.1 in the
          range 160$<$$D$$<$280.
    \item The bulk density of Borasisi is $2.1_{-0.59}^{+0.58}$ g cm$^{-3}$, which is higher
          (but within error bars) than other CKBOs of similar size. The bulk densities of % 2003 MW$_{12}$ 
          Varda and 2001 QC$_{298}$
          are $1.25_{-0.43}^{+0.40}$ g cm$^{-3}$ and $1.14_{-0.30}^{+0.34}$ g cm$^{-3}$, respectively.
          Our re-analysis of 
          four targets ($D$$<$400 km) has decreased their density estimates and they are
          mostly between 0.5 and 1 g cm$^{-3}$ implying high macroporosity.
   \end{enumerate}

\begin{acknowledgements}
We are grateful to Paul Hartogh for providing computational resources at Max-Planck-Institut f\"ur
Sonnensystemforschung, Germany.
      Part of this work was supported by the German
      \emph{DLR} project numbers 50 OR 1108, 50 OR 0903, 50 OR 0904 and 50OFO 0903. M.~Mommert acknowledges support
      trough the DFG Special Priority Program 1385.
      C. Kiss acknowledges the support of the Bolyai Research Fellowship of the Hungarian Academy of Sciences,
      the PECS 98073 contract of the Hungarian Space Office and the European Space Agency, and the K-104607
      grant of the Hungarian Research Fund (OTKA). A. P\'al acknowledges the support from the grant LP2012-31
      of the Hungarian Academy of Sciences.
      P. Santos-Sanz would like to acknowledge financial support by spanish grant AYA2011-30106-C02-01
      and by the Centre National de la Recherche Scientifique (CNRS).
      J.~Stansberry acknowledges support by NASA through an award issued by JPL/Caltech. R.~Duffard
      acknowledges financial support from the MICINN (contract Ram\'on y Cajal). 
      JLO acknowledges
      support from spanish grant AYA2011-30106-C02-01 and European FEDER funds.
      NP acknowledges funding by the Gemini-Conicyt Fund, allocated to
      the project N\textsuperscript{\underline{o}} 32120036.
\end{acknowledgements}

\appendix
\section{Debiasing of size distributions}
\label{debiasing}
\subsection{Targets}
In the debiasing we use those measured targets which are compatible with the orbital element space of
CFEPS synthetic objects. Due to different classification used in our observing program, one cold CKBO
(2001 QB$_{298}$) and two hot CKBOs from the inner belt (2003 UR$_{292}$ and 2002 XW$_{93}$) have been
excluded from the debiasing. To prevent possible contamination between cold/hot sub-populations we have also
excluded four hot CKBOs, whose inclinations are not far above the $i$=4.5 deg cut-off limit (Quaoar, Altjira,
2001 QD$_{298}$ and 2000 OK$_{67}$). Three measured targets have their semi-major axis within the 
gap
in CFEPS objects reserved for the 2:1 mean motion resonance with Neptune. To our knowledge these three targets
are not in resonance, therefore they are included. In total, 25 hot CKBOs from the inner and main belts are included in the debiasing
as well as 12 cold CKBOs.

\subsection{Magnitude conversion}
CFEPS uses $H_g$ values for their synthetic objects. If V-R color is known then $H_g$ can
be converted into $H_V$. We use the average color of cold classicals: V-R=0.63$\pm$0.09
(N=49) and hot classicals: V-R=0.51$\pm$0.14 (N=43) from the MBOSS-2 data base.
Lupton (2005) conversion formulas and conversion uncertainties
are (see Footnote 2 for reference):
\begin{equation}
V = g - 0.5784*(g - r) - 0.0038, \quad  \sigma = 0.0054
\end{equation}
\begin{equation}
R = r - 0.1837*(g - r) - 0.0971, \quad  \sigma = 0.0106.
\end{equation}
Using the average V-R color we get
\begin{equation}
\mathrm{cold:} \quad V=g-0.52, \quad  \sigma = 0.09   
\end{equation}
\begin{equation}
\mathrm{hot:} \quad V=g-0.40, \quad  \sigma = 0.14
\end{equation}

\subsection{PACS detection limit}
\label{PACSlimit}
Many of the measured cold CKBOs were very faint, the flux densities being $<$5.5 mJy.
Contrary, the hot CKBOs were brighter and only one out of 29 observed by PACS was a
non-detection.
The PACS observations were executed in a standardized way using similar observation durations and parameters.
While the repetition factor in ``TNOs are Cool'' was designed separately for each target, in the range REP=1,...,6,
for the cold CKBOs the most common choise was REP=5 (total 2-visit duration at 70 $\mu m$ or 100 $\mu m$ band 94 min).
Of the cold CKBOs only Sila, Teharonhiawako, 2001 XR254 and 2002 GV31 had shorter durations with REP=3
or REP=4. 2002 GV31 was a non-detection, the other three are large (D$>$200 km) and relatively bright ($H_V$$<$6.1).
The lowest flux densities with 2-band detections in our sample are at the level of 1.7 mJy at 100 $\mu m$.
According to NEATM (in the following we
assume $\eta=$1.2), the peak flux density of 1.7 mJy
would be emitted by an object at $r_h$=$\Delta$=40 AU
if its diameter is $D_\mathrm{ref}$=167.5 km and geometric albedo $p_{V,\mathrm{ref}}$=0.04. In the
following, we make the simplifying assumption that during our observations $r_h \approx \Delta$.
The peak flux density remains constant (but with a small shift in wavelength
position) if a target is at different distance and its size with respect to the reference size
is scaled according to the distance
change. If $s$ is a scale factor in the distance then the diameter scales as $s^{1.75}$, i.e. at a distance
of 40$s$ AU the object's size should be $s^{1.75} D_\mathrm{ref}$ to maintain the same peak flux density.
In the above, albedo was kept constant. The effect of albedo depends on the phase integral because we have
for the Bond albedo $A$:
$A=p_Vq\left(p_V\right)$, where $q\left(p_V\right)=0.336p_V+0.479$ (Brucker et al. 2009). Other values
being constant, if geometric albedo changes then the diameter has to be scaled in order to maintain the
constant maximum flux density. If $t$ is the scaling factor of geometric albedo, then diameter scales as
\begin{equation}
\left(\frac{1-p_{V,\mathrm{ref}}\,q(p_{V,\mathrm{ref}})}{1-t\, p_{V,\mathrm{ref}}\, q(t\, p_{V,\mathrm{ref}})}\right)^{\frac{3}{8}}.
\end{equation}

\subsection{Debiasing procedure}

\subsubsection{Debiasing stage 1} 
\label{stage1}
First we debias with respect to the radiometric
detection limit (Section~\ref{PACSlimit}).
Diameters are assigned to each CFEPS CKBO in a random way
using the geometric albedo probability densities derived from
measured targets (see Section~\ref{albedodiscussion}).
Then, size distributions of synthetic objects are calculated.
The debiasing factors of stage (1) are obtained by dividing the numbers of
CFEPS CKBOs
in the cumulative size distribution bins
by the numbers of potentially detectable 
CFEPS objects in the same size bins. 
For each synthetic object we have the distance, randomly selected albedo, and
the diameter calculated using that albedo and $H_g$. Each object is checked
against the detection limit derived in Section~\ref{PACSlimit}.
The uncertainties are calculated as the 1$\sigma$
uncertainties of the calculated ratio, where the two distributions have been produced 500
times with randomly assigned albedos to the synthetic objects.

The debiasing factors are applied to the size distributions of measured targets.
The numbers of targets in the measured size distribution bins are multiplied by
the corresponding stage (1) debiasing factors.

\subsubsection{Debiasing stage 2}
In stage (2) we are debiasing the selection effects of our target
sample compared to the sample which could have been detected with PACS.
The selection effects of the measured target sample include the
discovery bias of known TNOs.
This debiasing is done using optical absolute magnitudes $H_g$ of both the
synthetic CFEPS objects and the measured targets (steps 1-3 below) and translated into
debiasing factors for each bin in the size distribution (steps 4-6).
The stage (2) debiasing factors are
calculated in the following way:
\begin{enumerate}
 \item Create cumulative $H_g$ distribution of both the measured sample
       and the potentially detectable CFEPS synthetic objects. The
       latter is an average of a large number of sets of potentially detectable
       objects, where the detection limit calculation is using randomly assigned
       albedos (from the probability density distribution similar to those in
       Fig.~\ref{albedo_histo}) for each synthetic object.
 \item Calculate the ratio of numbers of objects in each $H_g$ bin of the
       potentially detectable distribution and the measured distribution.
       Normalize these factors so that the smallest factor is equal to one.
 \item Multiply the measured $H_g$ distribution by the factors from step 2.
 \item Generate sizes for objects in each $H_g$ bin
       after step 3 in a statistical
       way using the measured albedo probability density distribution.
       The relative differences in the numbers of objects in each $H_g$ bin of this step
       is given by the relative differences of numbers of objects in the $H_g$
       distribution of step 3.
 \item Calculate a size distribution using all the objects generated in step 4.
 \item Calculate debiasing factors from step 5 and the measured size distribution.
       Normalize these factors so that the largest target has a factor equal to one.
       In the dynamically hot sub-population Haumea and Makemake, two targets outside
       the CFEPS $H_g$ range, were not included in calculating debiasing factors.
\end{enumerate}

\end{document}